\documentclass[12pt,a4paper]{article}
\usepackage[left=1in,right=1in,top=1.25in,bottom=1.25in]{geometry}

\usepackage{mathrsfs}
\usepackage{amssymb}
\usepackage{amsmath}
\usepackage{ascmac}
\usepackage{amsthm}
\usepackage{bm}
\usepackage[dvips]{graphicx}
\usepackage{natbib}
\usepackage{setspace}
\usepackage{algorithm,algpseudocode}
\algnewcommand{\To}{\textbf{To }}
\usepackage{times}
\usepackage[colorlinks,linkcolor=blue,citecolor=blue,urlcolor=blue]{hyperref} 
\usepackage{placeins} 
\usepackage{authblk} 
\usepackage{color}
\usepackage{orcidlink}
\allowdisplaybreaks 
\makeatletter
\g@addto@macro\normalsize{%
  \setlength\abovedisplayskip{8pt}
  \setlength\belowdisplayskip{8pt}
  \setlength\abovedisplayshortskip{8pt}
  \setlength\belowdisplayshortskip{8pt}
}
\makeatother

\begin{document}
\title{Finite Mixtures of Generalized Estimating Equations for Clustering Multivariate Correlated Outcomes}
\author[1]{Shonosuke Sugasawa\orcidlink{0000-0002-9495-4280}\thanks{Corresponding author: sugasawa@econ.keio.ac.jp; Faculty of Economics, Keio University, Minato-ku, Tokyo, 108-8345, Japan}} 
\author[2]{Francis K.C. Hui\orcidlink{0000-0003-0765-3533}} 

\affil[1]{Faculty of Economics, Keio University, Tokyo, Japan}
\affil[2]{Research School of Finance, Actuarial Studies and Statistics, The Australian National University, Canberra, Australia}

\date{}
\maketitle
\thispagestyle{empty} 

\begin{abstract}
Multivariate correlated outcomes occur across disciplines, including ecology, social sciences, and psychometrics. This paper focuses on clustering these outcomes across observational units, specifically, finding groups of units with the same ``outcome profile". Our motivation comes from bioregionalization in ecology, which aims to cluster sites into bioregions with the same species profiles, where site membership can depend on environmental or habitat covariates. To accomplish this, we propose finite mixtures of generalized estimating equations (MixGEE). Unlike existing approaches to model-based bioregionalization, MixGEE partitions sites into regions while accounting for between-species correlations through a region-specific working correlation structure. Thus, each region is characterized by a marginal species mean vector and a between-species correlation matrix. Unlike likelihood-based finite mixture models, MixGEE does not require a full joint distribution of the multivariate outcomes. Instead, we construct a pseudo-posterior probability for region membership motivated by the large-sample distribution of the estimating equation. This leads to an iterative algorithm alternating between updating these probabilities and solving weighted estimating equations. We determine the number of regions using cross-validation based on predictive performance for held-out sites and species components, and use a clustered Dirichlet random-weight bootstrap for uncertainty quantification. Simulations demonstrate reliable estimation and inference under various correlation structures and more stable selection of the number of groups than methods that ignore dependence. Applying MixGEE to presence--absence records of fish species around the Kerguelen Plateau reveals three distinct fish assemblage profiles with heterogeneous occurrence and within-site correlation patterns.

\end{abstract}

\section{Introduction}
Multivariate correlated outcome data are regularly collected across many fields, from ecology and environmental studies to finance, psychometrics, and social studies \citep{song2007correlated,warton2015so}. This paper focuses on model-based clustering of multivariate outcomes across a set of observational units. That is, we aim to identify groups of units that have the same ``outcome profile", where the probability of a unit belonging to a profile is driven by covariates recorded at that site. Such a clustering task, or some variation thereof, is common in many statistical analyses, including grouping individuals into risk profiles based on insurance claims \citep[e.g.,][]{fung2019class,zhang2026spatially}, and disease subtyping when concomitant variables such as genetic markers are available \citep{li2024outcome}. This paper in particular is motivated by bioregionalization of multivariate abundance data in ecology \citep{hill2020determining, woolley2020bioregions}, which broadly speaking involves clustering sites into so-called bioregions (also commonly referred to as ecoregions, although we will also use the more generic term `region' throughout this paper) which have the same species profiles. Moreover, the probability of a site belonging to a region is driven by environmental and habitat factors, such that two sites possessing similar environmental covariates are more likely to belong to the same bioregion. Bioregionalization serves a number of important purposes in ecology, from identifying biodiversity hotspots with relatively unique species, to identifying regions likely to be strongly affected by climate change and develop targeted conservation and restoration strategies specific to those regions \citep{woolley2020bioregions,falaschi2023global}; see also Section \ref{sec:app} for a specific example of constructing marine bioregions based on presence-absence records of 20 fish species surveyed at 524 sites across the Kerguelen Plateau in the Indian Ocean southwest of Australia.

Within the statistical ecology literature, model-based approaches to bioregionalization remain underdeveloped and have largely concentrated on a basic form of the mixture of experts model \citep[Chapter 12,][]{fruhwirth2019handbook} where environmental covariates drive the so-called gating functions (only), and conditional on a site belonging to a region, species are assumed to be independent of each other \citep[see the region of common profiles or RCP model of][]{foster2013modelling,foster2017ecological}. While there has been some recent research extending this to handle spatial or spatiotemporal bioregionalization \citep{vanhatalo2021spatiotemporal}, to employ latent Dirichlet and mixed membership models instead \cite[e.g.,][]{valle2022latent}, and to construct more interpretable profiles in the case where the number of species is large \citep{zhou2025bayesian}, many if not all of these approaches do not explicitly account for (residual) between-species correlations that may be imperative to the characterization of the region itself, and/or require specification of a joint likelihood function across all species that can lead to computational challenges when it comes to estimation; see also \citet{warton2015so} and \citet{ovaskainen2020joint} among many others for more general work expounding the importance of accounting for between-species correlations when analyzing multivariate abundance data. We also acknowledge existing research on distance-based approaches to bioregionalization \citep[e.g.,][]{denelle2025bioregionalization}, as well as two-step approaches that perform bioregionalization in a fitted joint/stacked species distribution model \citep[e.g.,][]{ovaskainen2017make,lopez2026mapping}. The former lacks an explicit model formulation which may incur a variety of inferential and interpretability problems \citep{warton2015so, warton2017central}, while the latter often does not formally propagate the uncertainty from the fitted model to the clustering process itself.

Motivated by some of the above limitations, we propose a new finite mixture of generalized estimation equations, abbreviated MixGEE, for model-based bioregionalization of multivariate abundance data. As the name suggests, MixGEE involves formulating two sets of estimating equations: one for estimating the effects of environmental covariates driving the probabilities of region classification, along with a second set of region-specific weighted GEEs for estimating marginal mean vectors and a marginal working between species correlation matrix. The latter establishes two clear differences with the existing model-based bioregionalization techniques reviewed above, namely not having to specify a full likelihood function for each region and the estimation of a working correlation matrix to capture residual correlation between species. While the usual setup and application of GEEs in longitudinal data analysis (say) often treats the working correlation matrix as somewhat of a nuisance and only affects statistical efficiency and robustness properties \citep{zeger1988models,wang2014generalized}, in MixGEE it plays a more central role in the model, as each bioregion is defined by both its marginal species mean vector \emph{and} its correlation matrix between species. This provides an ecologically more realistic characterization of how species assemblages function as part of a bioregion, and, as we will show later in our application to the Kerguelen Plateau fish species data, can lead to noticeable differences in inference compared with RCP models which do not account for this; see \citet{Tho2027JointMeanCorrelation} and \citet{hui2025adjusted} for other developments of GEEs where the working correlation plays a central role in the analysis/inferential task. 

All the estimating equations underlying MixGEE require knowledge of the latent (unobserved) region membership indicators i.e., we need to know which region each site belongs to. To circumvent this problem, we construct a pseudo-posterior probability for region membership based on a measure of compatibility between each site and each region, and inspired by known large-sample distributional results for (weighted) GEEs \citep{zeger1988models,wang2011gee,wang2014generalized} along with related results concerning the associated score statistic \citep[e.g.,][]{rotnitzky1990hypothesis,hui2023gee}. This then leads to an iterative algorithm for fitting MixGEE not too dissimilar in spirit to that of an expectation-maximization algorithm used for maximum likelihood estimation for finite mixture and mixture of expert models \citep{fruhwirth2019handbook}. Specifically, we iterate between updating the pseudo-posterior probabilities given current parameter values, and solving the two sets of aforementioned estimating equations/GEEs to update covariate effects and parameters characterizing each bioregion, where the latent (unobserved) region membership indicators are replaced by the pseudo-posterior probabilities in the latter. Note as part of solving the GEE and to update the region-specific working correlation matrices, we assume an unstructured form and employ a regularized version of the sample correlation matrix based on the Pearson residuals \cite{warton2011regularized}. 
For inference, we use a clustered bootstrap to construct confidence intervals of the model parameters. 
Furthermore, to select the number of regions in MixGEE, we propose a $K$-fold cross-validation approach that evaluates out-of-sample classification and prediction performance based on simultaneously holding subsets of sites and species.

The proposed MixGEE is related to several strands of research in the broader statistical literature on model-based clustering for multivariate correlated outcomes. 
For example, using GEEs, \citet{tang2016mixture} developed a mixture modeling framework by modeling heterogeneity in the working correlation structure, while \citet{ito2023grouped} proposed grouped GEEs for heterogeneous longitudinal data under a hard-clustering formulation. More recently, \citet{liang2023multivariate} introduced multivariate mixture GEE models for longitudinal data with multiple outcomes by modeling heterogeneity in the working correlation structure, and \citet{liang2024heterogeneous} studied grouped finite mixture models for clustered heterogeneous data. 
To the best of our knowledge though, this paper is the first to adopt such ideas of finite mixtures of GEEs to the statistical ecology literature, let alone for bioregionalization of multivariate abundance data. 
As discussed earlier, there also exists a variety of likelihood-based mixture modeling approaches to clustering multivariate correlated outcome data, though they often involve positing a hierarchical/joint distribution for all the outcomes \citep[see also][]{jara2007dirichlet,cagnone2012factor,scharf2022multivariate,stratton2024clustering}. 
There remains a need for a computationally light alternative for finite mixture modeling of correlated multivariate outcomes, in which between-outcome correlation (and subsequently cluster assignment) can be accounted for without constructing a parametric joint distribution for the responses. 
Simulations show that the proposed MixGEE accurately recovers the latent group structure and provides reliable estimation and inference across correlation structures.  
Applying MixGEE to perform model-based bioregionalization of the Kerguelen Plateau fish species data reveals three distinct assemblage profiles and substantial heterogeneity in the within-site correlation patterns across regions.

The remainder of the paper is structured as follows. Section \ref{sec:mixgee} formulates MixGEE, including construction of the pseudo-posterior probabilities of region membership, along with details regarding uncertainty quantification and for selecting the number of regions via cross-validation. Section \ref{sec:sim} presents results of a simulation study comparing a number of variations of MixGEE to the RCP models, while Section \ref{sec:app} applies MixGEE for bioregionalization of the Kerguelen Plateau fish species data. 
Finally, Section~\ref{sec:conc} offers some concluding remarks and avenues for future research.

\section{Finite mixture of generalized estimating equations} \label{sec:mixgee}
Consider a set of $n$ independent observational units (e.g., sites), where at unit $i=1,\ldots,n$, we observe a set of $p$-vector of responses (e.g., species) $\bm{y}_i = (y_{i1},\ldots,y_{ip})^\top$, along with a $q$-vector of covariates $\bm{x}_i$. We assume the first element of $\bm{x}_i$ is set to one to represent an intercept term. Note the covariates are assumed to be common across all responses, although the developments below can be straightforwardly extended to accommodate unit-and-response-specific covariates. 

To perform model-based clustering, we assume each of the $n$ observational sites belongs to one of $G \ll n$ regions (clusters), such that all units in the same regions are characterized by the same set of parameters. Analogous to mixture of expert models, the probability of region membership is then modeled as a function of covariate. In detail, let $z_{ig}$ denote the latent (unobserved) region membership label for site $i$, where $z_{ig} = 1$ if site $i$ belongs to region $g$ and zero otherwise, and $\sum_{g=1}^G z_{ig} = 1$. Then we model the (prior) probability of membership via a multinomial logistic-type setup: 
\begin{align*}
\mathbb{P}(z_{ig} = 1) = \pi_{ig}(\bm{\alpha}) \equiv 
\frac{\exp(\bm{x}_i^\top\bm{\alpha}_g)}
{\sum_{g'=1}^G \exp(\bm{x}_i^\top\bm{\alpha}_{g'})},
\end{align*}
for $g = 1,\ldots,G$, where $\bm{\alpha}_g$ denotes the $q$-vector of regression coefficients corresponding to the probability of belonging to region $g$, and $\bm{\alpha} = (\bm{\alpha}_1^\top, \ldots, \bm{\alpha}_G)^\top$. As is typical in many multinomial regression setups \citep[e.g.,][]{agresti2012categorical}, we constrain one of the coefficient vectors to zero for reasons of parameter identifiability; in particular, here we set $\bm{\alpha}_1 = \bm{0}_q$. To estimate $\bm{\alpha}$, we use the following set of estimation equations. Given the full $nG$-vector of region membership labels, $\bm{z} = (z_{11},\ldots,z_{1G},z_{21},\ldots,z_{nG})^\top$, we solve
$\mathcal{S}_{\bm{\alpha}}(\bm{z}; \bm{\alpha}) = \{\mathcal{S}_{\bm{\alpha},2}(\bm{z}; \bm{\alpha}_2)^\top, \ldots, \mathcal{S}_{\bm{\alpha},G}(\bm{z}; \bm{\alpha}_G)\}^\top = \bm{0}_{(G-1)q}$, where for $g = 2,\ldots,G$,
\begin{align}
\mathcal{S}_{\bm{\alpha},g}(\bm{z}; \bm{\alpha}_g) = \sum_{i=1}^n \bm{x}_i\left\{z_{ig} - \pi_{ig}(\bm{\alpha})\right\}.
\label{eq:EE-z}
\end{align}
It is straightforward to see that the above estimating equations are unbiased, since $\mathbb{E}(z_{ig}|\bm{x}_i)=\pi_{ig}(\bm{\alpha})$, and hence $\mathbb{E}[\mathcal{S}_{\bm{\alpha},g}(\bm{z};\bm{\alpha}_g)|\bm{X}]=\bm{0}_q$ at the true parameter value.

Next, within region $g$, we assume the responses have marginal mean and variance given by $\mathbb{E}[y_{ij} | z_i = g] = \mu_{gj} = h(\eta_{gj})$ and ${\rm Var}[y_{ij} | z_i = g] = \gamma_{gj} v\{h(\eta_{gj})\}$, where $h(\cdot)$ and $v(\cdot)$ denote the assumed inverse link and variance functions, respectively, and $\eta_{gj}$ and $\gamma_{gj} > 0$ denote linear predictor and dispersion parameters, respectively, for response $j$ in region $g$. Examples of commonly chosen inverse link and variance functions (and dispersion parameters) include $h(u) = \{1+\exp(u)\}^{-1}\exp(u)$, $v(u) = u(1-u)$, and $\gamma_{gj} = 1$ in the case of binary responses, $h(u) = \exp(u)$, $v(u) = u$, and $\gamma_{gj} = 1$ in the case of count responses, and $h(u) = \exp(u)$ and $v(u) = u^2$ in the case of positive continuous responses; see also Appendix \ref{app:mixgeedetails} for further details in the case of the first two response types, noting in both these cases the dispersion parameters are fixed \emph{a priori}. The developments below can also be extended to the case where we assume a more general form for the marginal variance e.g., $v\{h(\eta_{gj})\}$ itself depends on one or more unknown dispersion parameters as in the case of quadratic or power-law variance functions characteristic of the negative binomial and Tweedie distributions \citep{hui2022gee, stoklosa2022overview}, although for simplicity we will use the multiplicative form set out above. More importantly, we highlight that the $\eta_{gj}$'s are specific to each region, but do not depend on the $\bm{x}_i$'s. That is, covariates only drive the probabilities of region membership, and not the ``profiles" of the regions themselves. 

Let $\bm{\eta}_g = (\eta_{g1},\ldots,\eta_{gp})^\top$ and $\bm{\eta} = (\bm{\eta}_1^\top, \ldots, \bm{\eta}_G^\top)^\top$ denote the region-specific and full vector of linear predictors, respectively, and analogously define $\bm{\gamma}_g = (\gamma_{g1},\ldots,\gamma_{gp})^\top$ and $\bm{\gamma} = (\bm{\gamma}_1^\top, \ldots, \bm{\gamma}_G^\top)^\top$. To estimate the former, and given $\bm{z}$, we solve a second set of region-specific generalized estimating equations or GEEs  $\mathcal{S}_{\bm{\eta}}(\bm{y}, \bm{z}; \bm{\eta}) = \{\mathcal{S}_{\bm{\eta},2}(\bm{y}, \bm{z}; \bm{\eta}_2)^\top, \ldots, \mathcal{S}_{\bm{\eta},G}(\bm{y}, \bm{z}; \bm{\eta}_G)\}^\top = \bm{0}_{Gp}$ and, for $g = 1,\ldots,G$,
\begin{align}\label{eq:EE-y}
\mathcal{S}_{\bm{\eta},g}(\bm{y}, \bm{z}; \bm{\eta}_g) = \sum_{i=1}^n z_{ig} \bm{S}_{\bm{\eta},g}(\bm{y}_i; \bm{\eta}_g) \equiv \sum_{i=1}^n z_{ig} \bm{D}_g^\top \bm{V}_g^{-1}(\bm{y}_i - \bm{\mu}_g),
\end{align} 
where $\bm{\mu}_g = (\mu_{g1},\ldots,\mu_{gp})^\top$, 
$\bm{D}_g = \partial\bm{\mu}_g/\partial\bm{\eta}_g = \text{Diag}\{h'(\eta_{g1}), \ldots, h'(\eta_{gp})\}$ is a $p \times p$ diagonal matrix, and $\bm{V}_g = \bm{A}_g^{1/2} \bm{R}_g \bm{A}_g^{1/2}$ denotes the working covariance matrix for the region $g$ with $\bm{A}_g = \text{Diag}[\gamma_{g1} v\{h(\eta_{g1})\},\ldots,\gamma_{gp} v\{h(\eta_{gp})\}]$, and $\bm{R}_g$ is the assumed working correlation matrix corresponding describing the (residual) covariation among the $p$ species. Given $\bm{z}$, it is clear that \eqref{eq:EE-y} is also unbiased. Moreover, this is unaffected by the specification of the working correlation $\bm{R}_g$, and indeed such a property forms the basis for establishing the large-sample consistency and asymptotic normality of GEE estimates and their robustness to misspecification of the working correlation \citep{zeger1988models,wang2011gee,wang2014generalized}.

On the other hand, the specification of the working correlation matrix typically affects the efficiency of statistical inference on $\bm{\eta}_g$ \citep{zeger1988models,wang2014generalized}, so choosing a reasonable form to capture the correlations of responses within a region remains important. Unlike longitudinal and spatial/spatio-temporal settings, however, without additional information there is no simple concept of distance between responses. As such, in this paper our default preference is to use a regularized form of an unstructured working correlation matrix estimate, specifically, $\bm{R}_g = (1 -\lambda) \widehat{\bm{R}}_g + \lambda \bm{I}_p$, where $\widehat{\bm{R}}_g = n_g^{-1} \sum_{i=1}^{n} z_{ig} \bm{r}_{ig}\bm{r}_{ig}^\top$ is the sample working correlation matrix computed based on Pearson residuals $\bm{r}_{ig} = \bm{A}_{g}^{-1/2} (\bm{y}_i - \bm{\mu}_g); i = 1,\ldots,n$, $n_g = \sum_{i=1}^n z_{ig}$, $\bm{I}_p$ denotes a $p \times p$ identity matrix, and $\lambda \in [0,1]$ controls the degree of shrinkage toward the independence working correlation. While $\lambda = 0$ could be used, the inclusion of shrinkage generally improves numerical stability when inverting the resulting working covariance matrix, particularly if $p$ is non-negligible compared to $n_g$. Throughout the numerical studies in this paper, we set $\lambda = 0.2$, but acknowledge that future research could use a data-driven approach to select $\lambda$ such as cross-validation \citep[e.g.,][]{warton2008penalized}; see also \citet{pan2001robust}, \citet{warton2011regularized} and \citet{wang2016covariance} for examples of other approaches to constructing the empirical working correlation matrix $\widehat{\bm{R}}_g$, \citet{bonat2016multivariate} and \citet{fan2024covariance} for other flexible approaches to constructing the working correlation matrices, and Section \ref{sec:conc} for discussion on assuming the same working correlation matrix across regions.
Note also $\lambda = 1$ and $\bm{R}_g = \bm{I}_p$ corresponds to the case of independence estimation equations (IEE), and while we anticipate these also perform well in terms of point estimation of $\bm{\eta}_g$, we expect them to be less statistically efficient in general.  

For the remainder of this paper, we refer to equations \eqref{eq:EE-z} and \eqref{eq:EE-y} as a finite mixture of generalized estimating equations, or MixGEE. Together, they achieve model-based clustering of observational units into regions (characterized by $\bm{\alpha}$), and define the profiles of the regions (characterized by $\bm{\eta}_g$ and the $\bm{R}_g$). The dispersion parameters for each region, $\bm{\gamma}_g$, are regarded as nuisance parameters and are either known \emph{a priori} or estimated as described in the next section.

\subsection{Estimation} \label{subsec:fitting}
If the region membership labels $\bm{z}$ were observed, then estimates of $\bm{\alpha}$, $\bm{\eta}$, and the working correlation matrices $\bm{R}_g$ (and indeed the dispersion parameters $\bm{\gamma}$ as appropriate) could be obtained by iterating between the estimating equations and updates described in the previous section. The main difficulty therefore lies in inferring the unknown region membership labels. 
To achieve this within the MixGEE framework, noting that there is no fully specified likelihood-based mixture function to make use of here, we propose to construct a so-called ``pseudo-posterior" distribution for the $z_{ig}$'s motivated by existing large-sample distributional results for independent cluster GEEs as follows. For region $g$, given $\bm{z}$ and under standard regularity conditions as $n_g \rightarrow \infty$ for fixed $p$ and $q$, we can adapt results from \cite{zeger1988models,wang2011gee,hui2023gee} among others and show that equation \eqref{eq:EE-y} satisfies
\begin{align}\label{eq:score-dist}
\frac{1}{n^{1/2}_g}\mathcal{S}_{\bm{\eta},g}(\bm{y}, \bm{z}; \bm{\eta}_g) = \frac{1}{n^{1/2}_g} \sum_{i=1}^n z_{ig} \bm{S}_{\bm{\eta},g}(\bm{y}_i;\bm{\eta}_g)
\xrightarrow{d} \mathcal{N}_p\left(\bm{0}_p, \bm{\Omega}_g\right),
\end{align}
at the true parameter values, where $\bm{\Omega}_g = \bm{D}_g^\top \bm{V}_g^{-1}\bm{V}_g^\ast \bm{V}_g^{-1}\bm{D}_g$ is the sandwich covariance matrix for region $g$ and $\bm{V}_g^\ast = {\rm Cov}[\bm{y}_i| z_{ig} = 1]$ denotes the true marginal covariance matrix.
Note while this limiting result formally holds only at the true parameter value, in this article it serves as motivation for a Gaussian working approximation evaluated at a candidate value $\bm{\eta}_g$ as seen shortly. 
Also, if the working marginal covariance matrix coincided with the true marginal covariance, then this expression would simplify to $\bm{\Omega}_g = \bm{D}_g^\top \bm{V}_g^{-1}\bm{D}_g$, that is, the so-called naive or model-based covariance matrix \citep{wang2014generalized}.
In practice, given suitable estimates of $\bm{\eta}_g$ then $\bm{V}_g^\ast$ can be estimated empirically by the sample covariance matrix $n_g^{-1}\sum_{i=1}^n z_{ig}(\bm{y}_i - \bm{\mu}_g)(\bm{y}_i - \bm{\mu}_g)^\top$, although there exists a number of variations for this and the sandwich covariance matrix more broadly in the literature, including ones based in sample working correlation $\widehat{\bm{R}}_g$ discussed earlier \citep[e.g.,][]{pan2001robust,warton2011regularized,wang2016covariance}.

Equation \eqref{eq:score-dist} suggests that, for each unit $i$ belonging to region $g$, their contribution to the overall GEE should be concentrated around zero with covariance determined by $\bm{\Omega}_g$. 
In particular, it motivates using the quadratic loss function $-2^{-1} \bm{S}_{\bm{\eta},g}(\bm{y}_i;\bm{\eta}_g)^\top \bm{\Omega}_g^{-1} \bm{S}_{\bm{\eta},g}(\bm{y}_i;\bm{\eta}_g)$, and subsequently $\phi_p\{\bm{S}_{\bm{\eta},g}(\bm{y}_i;\bm{\eta}_g) | \bm{0}_p,\bm{\Omega}_g\}$ where $\phi_p(\cdot | \bm{\mu},\bm{\Sigma})$ denotes the probability density function of a $p$-dimensional multivariate normal distribution with mean vector $\bm{\mu}$ and covariance matrix $\bm{\Sigma}$, as a measure of the compatibility between observation $i$ and region $g$. 
Combining this with the prior probability, we thus obtain the pseudo-posterior probability
\begin{align}\label{eq:pos-prob}
\tau_{ig} 
\equiv  \frac{\pi_{ig}(\bm{\alpha})\,
\phi_p\{\bm{S}_{\bm{\eta},g}(\bm{y}_i;\bm{\eta}_g) | \bm{0}_p,\bm{\Omega}_g\}}
{\sum_{g' = 1}^G
\pi_{ig'}(\bm{\alpha})\, 
\phi_p\{\bm{S}_{\bm{\eta},g'}(\bm{y}_i;\bm{\theta}_{g'}) | \bm{0}_p,\bm{\Omega}_{g'}\}},
\end{align}
for $i = 1,\ldots,n$ and $g = 1,\ldots,G$, where $\sum_{g=1}^G \tau_{ig} = 1$ by definition. 
Intuitively, if unit $i$ does not belong to region $g$, then its estimating-function contribution evaluated under region $g$ will generally be shifted away from zero, resulting in a smaller compatibility measure and hence a lower pseudo-posterior probability.
Note the similarity between \eqref{eq:pos-prob} and the form of posterior probabilities commonly seen in EM-type algorithms for maximum likelihood estimation of finite mixture models \citep{fruhwirth2019handbook}. In particular, $\tau_{ig}$ plays the role of soft-clustering the observational units into regions: it combines the prior membership probability $\pi_{ig}(\bm{\alpha})$ with the Gaussian working density of the estimating function. Equation \eqref{eq:pos-prob} also bears similarity in form to quadratic score statistics based on standardized estimating functions \citep{lindsay2003inference}, along with the use of the score statistics for variable selection in GEEs \citep[e.g.,][]{stoklosa2014fast,hui2022gee}.

Based on this pseudo-posterior formulation, we can now formally describe our approach to estimating MixGEE. Let $\bm{\tau}_i =(\tau_{i1},\ldots,\tau_{iG})^\top$, $\bm{\tau} = (\bm{\tau}_1^\top, \ldots, \bm{\tau}_n^\top)^\top$, and define the full parameter vector of interest as $\bm{\theta} = \{\bm{\alpha}^\top,\bm{\eta}^\top, \text{vech}(\bm{R}_1)^\top, \ldots, \text{vech}(\bm{R}_G)^\top, \bm{\gamma}^\top\}^\top$. Then at iteration $t = 0, 1, 2, \ldots,$ and given current parameter values $\bm{\theta}^{(t)}$, we first update the pseudo-posterior probabilities $\tau_{ig}^{(t+1)}$ based on \eqref{eq:pos-prob}. 
Next, we update $\bm{\alpha}^{(t+1)}$ by taking one blockwise
Newton--Raphson step toward solving \eqref{eq:EE-z}, with
$\bm{z}$ replaced by $\bm{\tau}^{(t+1)}$.
That is, we solve $\mathcal{S}_{\bm{\alpha}}(\bm{\tau}^{(t+1)}; \bm{\alpha}) = \bm{0}_{(G-1)q}$ with $\mathcal{S}_{\bm{\alpha},g}(\bm{\tau}^{(t+ 1)}; \bm{\alpha}_g) = \sum_{i=1}^n \bm{x}_i \{\tau^{(t+1)}_{ig} - \pi_{ig}(\bm{\alpha})\}$. This can be done using, say, a standard Newton--Raphson update, the details of which are provided in Algorithm \ref{algo:EEE} given below.
Finally, we update $\bm{\eta}^{(t+1)}$ and the working correlation matrices $\{\bm{R}_g^{(t+1)}; g = 1,\ldots,G\}$ by solving \eqref{eq:EE-y}, where again $\bm{z}$ is replaced by $\hat{\bm{\tau}}^{(t+1)}$, and using the regularized form of the unstructured working correlation matrix estimate discussed in the previous section. That is, we solve $\mathcal{S}_{\bm{\eta}}(\bm{y}, \bm{\tau}^{(t+1)}; \bm{\eta}) = \bm{0}_{Gp}$ with $\mathcal{S}_{\bm{\eta},g}(\bm{y}, \bm{\tau}^{(t+1)}; \bm{\eta}_g) = \sum_{i=1}^n \tau^{(t+1)}_{ig} \bm{D}_g^\top (\bm{V}^{(t)}_g)^{-1}(\bm{y}_i - \bm{\mu}_g)$ and $\bm{V}^{(t)}_g = (\bm{A}^{(t)}_g)^{1/2} \bm{R}^{(t)}_g (\bm{A}^{(t)}_g)^{1/2}$, along with $\bm{R}^{(t+1)}_g = (1 - \lambda)\bm{\Delta}_g^{-1/2}\widehat{\bm{R}}^{(t+1)}_g\bm{\Delta}_g^{-1/2} + \lambda \bm{I}_p$, where $\widehat{\bm{R}}^{(t+1)}_g=(n^{(t+1)}_g)^{-1} \sum_{i=1}^{n} \tau^{(t+1)}_{ig} \bm{r}^{(t+1)}_{ig}\bm{r}^{(t+1)\top}_{ig}$, $\bm{\Delta}_g=\operatorname{Diag}(\widehat{\bm{R}}^{(t+1)}_g)$ and $n^{(t+1)}_g = \sum_{i=1}^n \tau^{(t+1)}_{ig}$. 
If necessary, we can also update the dispersion parameter  $\bm{\gamma}^{(t+1)}$ using a simple moment-based estimator i.e., for $g = 1,\ldots,G$ and $j = 1,\ldots, p$, we set $\gamma_{gj}^{(t+1)} = [n^{(t+1)}_g v\{h(\eta^{(t+1)}_{gj})\}]^{-1} \sum_{i=1}^n \tau^{(t+1)}_{ig} (y_{ij} - \mu^{(t+1)}_{gj})^2$. 

We iterate between the above steps until convergence, which is assessed
using $\bm{\theta}_{\mathrm{mon}}=(\bm{\eta}^{\top},\bm{\alpha}^{\top},\bm{\gamma}^{\top})^{\top}$, excluding the working correlation matrices. 
Specifically, the algorithm is terminated when $\|\bm{\theta}_{\mathrm{mon}}^{(t+1)}-\bm{\theta}_{\mathrm{mon}}^{(t)}\|<\epsilon$ for some small tolerance value $\epsilon>0$.
Algorithm \ref{algo:EEE} summarizes the estimation procedure for MixGEE, noting step-halving can also be employed in the Newton-Raphson updates as appropriate.
For computational efficiency, we perform a single Newton update of $\bm{\alpha}$ at each iteration rather than fully solving its estimating equation conditional on the current pseudo-posterior probabilities.
Thus, $\bm{\alpha}$ is updated progressively over the outer iterations together with the other model components.

\begin{algorithm}[htb]
\caption{Estimation of MixGEE.} \label{algo:EEE}
\begin{algorithmic}
\Require Multivariate correlated data $\{(\bm{x}_i, \bm{y}_i); i = 1,\ldots,n\}$, initial parameter values $\bm{\theta}^{(0)} = \{\bm{\alpha}^{(0)\top},\bm{\eta}^{(0)\top}, \text{vech}(\bm{R}_1)^{(0)\top}, \ldots, \text{vech}(\bm{R}_G)^{(0)\top}, \bm{\gamma}^{(0)\top}\}^\top$, regularization parameter $\lambda \in [0,1]$, tolerance value $\epsilon > 0$. 

\Repeat $\quad t = 0,1,2\ldots$
\begin{description}
\item[i.] 
Update the pseudo-probability $\tau_{ig}^{(t+1)}$ using \eqref{eq:pos-prob}.

\item[ii.] 
Update $\bm{\alpha}^{(t+1)}$ by taking one blockwise Newton--Raphson step toward solving $\mathcal{S}_{\bm{\alpha}}(\bm{\tau}^{(t+1)}; \bm{\alpha}) = \bm{0}_{(G-1)q}$. 
That is, for $g = 2, \ldots, G$, 
\begin{align*}
\bm{\alpha}_g^{(t+1)} = \bm{\alpha}_g^{(t)} +
\left[ \sum_{i=1}^n \pi_{ig}(\bm{\alpha}^{(t)}) \left\{1 - \pi_{ig}(\bm{\alpha}^{(t)}) \right\} \bm{x}_i\bm{x}_i^\top \right]^{-1}
\sum_{i=1}^n \left\{\tau_{ig}^{(t+1)} - \pi_{ig}(\bm{\alpha}^{(t)}) \right\} \bm{x}_i.    
\end{align*}

\item[iii.] 
For $g=1,\ldots,G$, update $\bm{\eta}^{(t+1)}_g$ as
\[
\bm{\eta}^{(t+1)}_g
=
h^{-1}\left\{
\frac{1}{n_g^{(t+1)}}
\sum_{i=1}^n \tau_{ig}^{(t+1)}\bm{y}_i
\right\},
\qquad
n_g^{(t+1)}
=
\sum_{i=1}^n\tau_{ig}^{(t+1)},
\]
where $h^{-1}$ is applied componentwise.

\item[iv.] 
For $g=1,\ldots,G$, update the working correlation matrix as $\widehat{\bm{R}}^{(t+1)}_g = (n^{(t+1)}_g)^{-1} \sum_{i=1}^{n} \tau^{(t+1)}_{ig} \bm{r}^{(t+1)}_{ig}\bm{r}^{(t+1)\top}_{ig}$ and $\bm{R}^{(t+1)}_g = (1 - \lambda)\bm{\Delta}_g^{-1/2}\widehat{\bm{R}}^{(t+1)}_g\bm{\Delta}_g^{-1/2} + \lambda \bm{I}_p$ with $\bm{\Delta}_g=\operatorname{Diag}(\widehat{\bm{R}}^{(t+1)}_g)$. 

\item[v.]
If required, for $g = 1,\ldots,G$ and $j = 1,\ldots,p$, update the dispersion parameters as
\begin{align*}
\gamma_{gj}^{(t+1)} = \frac{1}{n^{(t+1)}_g} \sum_{i=1}^n \tau^{(t+1)}_{ig} \frac{(y_{ij} - \mu^{(t+1)}_{gj})^2}{v\{h(\eta^{(t+1)}_{gj})\}]}.    
\end{align*}
\end{description}

\Until{ Convergence criterion met e.g.,  $\|\bm{\theta}_{\mathrm{mon}}^{(t+1)}-\bm{\theta}_{\mathrm{mon}}^{(t)}\|\leq\epsilon$, where $\bm{\theta}_{\mathrm{mon}}=(\bm{\eta}^{\top},\bm{\alpha}^{\top},\bm{\gamma}^{\top})^{\top}$.}
\end{algorithmic}
\end{algorithm}

It is not hard to see that Algorithm \ref{algo:EEE} resembles an EM-type procedure for fitting finite mixture and mixture of experts models, where step i corresponds to the E-step and steps ii to v correspond to effectively a set of conditional M-steps. 
As starting values, we set $\bm{\alpha}^{(0)}=\bm{0}$, corresponding to equal prior membership probabilities, and obtain an initial partition by applying Ward's hierarchical clustering to the Euclidean distances between the response vectors $\bm{y}_i$. 
The initial $\bm{\tau}^{(0)}$ is then given by the resulting hard cluster assignments, while each $\bm{\mu}_g^{(0)}$ is set to the within-cluster sample mean and $\bm{\eta}_g^{(0)}=h^{-1}(\bm{\mu}_g^{(0)})$. 
The initial working correlation matrices are computed from the corresponding Pearson residuals using the regularized unstructured estimator described above.
Any dispersion parameters are initialized at one.
The proposed algorithm is computationally efficient, as the region-specific means and working correlation matrices admit closed-form updates and the membership parameters require only Newton updates. 
The specific computation time will be provided in Section~\ref{sec:app}.
For the remainder of the paper, we let $\hat{\bm{\theta}} = \{\hat{\bm{\alpha}},\hat{\bm{\eta}}, \text{vech}(\hat{\bm{R}}_1)^{\top}, \ldots, \text{vech}(\hat{\bm{R}}_G)^{\top}, \hat{\bm{\gamma}}^{\top}\}^\top$ denote the MixGEE estimates based on applying Algorithm \ref{algo:EEE}, and $\hat{\bm{\tau}}$ the corresponding pseudo-posterior probabilities at convergence. 

\subsection{Standard errors and inference}\label{sec:variance}

The pseudo-posterior probability in (\ref{eq:pos-prob}) relies on the covariance structure of the score statistics derived from the estimating equations. 
In this subsection we describe how these covariance matrices can be estimated and how they lead to asymptotic inference for the parameter estimators.

First consider the estimating equation for $\bm{\alpha}$ in (\ref{eq:EE-z}). 
Let $\psi_{ig}(\bm{\alpha})=\{\tau_{ig}-\pi_{ig}(\bm{\alpha})\}\bm{x}_i$ denote the score contribution from observation $i$. 
Under standard regularity conditions for multinomial logistic models, the asymptotic covariance matrix of the score function can be estimated by the empirical covariance of these contributions. 
Hence, given $\bm{\alpha}$, the asymptotic variance of the score function in (\ref{eq:EE-z}) can be estimated by $n^{-1}\sum_{i=1}^n \{\tau_{ig}-\pi_{ig}(\bm{\alpha})\}^2 \bm{x}_i\bm{x}_i^\top$.
Next consider the estimating equation for the response parameters in (\ref{eq:EE-y}). 
Let $\mathcal{S}(\bm{y}_i;\bm{\eta}_g)$ denote the individual score contribution for cluster $g$. 
According to the asymptotic theory of generalized estimating equations, the aggregated score satisfies the distribution given in (\ref{eq:score-dist}). 

Let $\mathcal{H}_g = \mathbb{E}[-\partial \mathcal{S}(\bm{y}_i;\bm{\eta}_g)/\partial \bm{\eta}_g^\top]$ denote the sensitivity matrix and 
$\mathcal{J}_g = {\rm Cov}\{\mathcal{S}(\bm{y}_i;\bm{\eta}_g)\}$ denote the variability matrix. 
Then the asymptotic covariance matrix of the estimator $\widehat{\bm{\eta}}_g$ is given by the sandwich form $\bm{\Sigma}_{\bm{\eta}_g}=\mathcal{H}_g^{-1}\mathcal{J}_g\mathcal{H}_g^{-1}$.
In practice, $\mathcal{J}_g$ can be estimated empirically using the covariance of the score contributions. 
Accordingly, given $\bm{\theta}$ and $\bm{\alpha}$, the asymptotic variance of the score function in (\ref{eq:EE-y}) can be estimated by $n^{-1}\sum_{i=1}^n \tau_{ig}^2 \mathcal{S}(\bm{y}_i;\bm{\eta}_g)\mathcal{S}(\bm{y}_i;\bm{\eta}_g)^\top$.
Similarly, the asymptotic covariance matrix of the estimator $\widehat{\bm{\alpha}}_{-1}\equiv(\widehat{\bm{\alpha}}_2^\top,\ldots,\widehat{\bm{\alpha}}_G^\top)^\top$ can be expressed as $\bm{\Sigma}_{\bm{\alpha}}=\mathcal{H}_{\bm{\alpha}}^{-1}\mathcal{J}_{\bm{\alpha}}\mathcal{H}_{\bm{\alpha}}^{-\top}$, where $\mathcal{H}_{\bm{\alpha}}= \mathbb{E}\left(\left[\operatorname{Diag}\{\bm{\pi}_{i,-1}(\bm{\alpha})\}-\bm{\pi}_{i,-1}(\bm{\alpha})\bm{\pi}_{i,-1}(\bm{\alpha})^\top\right]\otimes\bm{x}_i\bm{x}_i^\top\right)$, $\mathcal{J}_{\bm{\alpha}} = {\rm Cov}\left[\{\bm{\tau}_{i,-1}-\bm{\pi}_{i,-1}(\bm{\alpha})\}\otimes\bm{x}_i\right]$ and $\bm{\pi}_{i,-1}(\bm{\alpha})=\{\pi_{i2}(\bm{\alpha}),\ldots,\pi_{iG}(\bm{\alpha})\}^\top$, with $\bm{\tau}_{i,-1}$ defined analogously.
The matrices $\mathcal{H}_{\bm{\alpha}}$, $\mathcal{J}_{\bm{\alpha}}$, $\mathcal{H}_g$, and $\mathcal{J}_g$ can be estimated by their empirical counterparts obtained from the fitted model. 
Substituting these estimates yields asymptotic covariance estimators for $\widehat{\bm{\alpha}}_{-1}$ and $\widehat{\bm{\theta}}_g$, which provide standard errors for the parameter estimates and enable Wald-type confidence intervals.
When the working correlation structure is correctly specified, the model-based covariance estimator may be used, which replaces $\mathcal{J}_g$ with $\mathcal{H}_g$. 
Otherwise, the sandwich estimator provides robustness against misspecification of the correlation structure.

A potential limitation of the above asymptotic variance formula is that it does not account for the additional variability arising from estimation of the working correlation matrices. To better capture this source of uncertainty, we employ a clustered random-weight bootstrap. 
For bootstrap replication $b=1,\ldots,B$, we generate $(w_1^{(b)},\ldots,w_n^{(b)})^\top\sim n\mathrm{Dirichlet}(1,\ldots,1)$ and refit the entire MixGEE algorithm after replacing each sum over observational units by its weighted counterpart. 
Within each bootstrap replication, the pseudo-probabilities are updated using \eqref{eq:pos-prob} with the current bootstrap-specific parameter estimates. 
The random weights do not enter the pseudo-probability formula directly, but affect the pseudo-probabilities indirectly through the weighted updates of the model parameters. 
Specifically, in bootstrap replication $b$, the estimating equations for $\bm{\alpha}_g$ and $\bm{\eta}_g$ are replaced by
\begin{align*}
\mathcal{S}_{\bm{\alpha},g}^{(b)}(\bm{\tau};\bm{\alpha}_g)
&=\sum_{i=1}^n w_i^{(b)}\bm{x}_i\{\tau_{ig}-\pi_{ig}(\bm{\alpha})\}, \quad 
\mathcal{S}_{\bm{\eta},g}^{(b)}(\bm{y},\bm{\tau};\bm{\eta}_g)
=\sum_{i=1}^n w_i^{(b)}\tau_{ig}\bm{S}_{\bm{\eta},g}(\bm{y}_i;\bm{\eta}_g),
\end{align*}
and the working correlation matrices are updated using the same weighted estimating procedure.
The same weight $w_i^{(b)}$ is applied to all components of $\bm{y}_i$, thereby preserving the within-unit dependence structure. 
Because the working correlation matrices and all other model components are re-estimated in each bootstrap replication, the variability of the resulting estimates incorporates uncertainty arising from the full estimation procedure.
Since the group labels may be permuted across bootstrap fits, we align
the labels in each bootstrap replication with those from the original
fit by selecting the permutation that minimizes the squared Euclidean
distance between the corresponding estimates of $\bm{\eta}$.
After label alignment, standard errors are estimated by the empirical
standard deviations of the bootstrap estimates, and pointwise
Wald-type confidence intervals are constructed using these bootstrap
standard errors.

\subsection{Selection of the number of groups}
\label{sec:CV}

To select the number of groups $G$, we employ a cross-validation procedure based on predictive performance.
Let $\{1,\ldots,n\}$ denote the set of sites and $\{1,\ldots,p\}$ denote the set of response components.
We partition the sites into $K$ subsets $I_1,\ldots,I_K$ and the response components into $K$ subsets $J_1,\ldots,J_K$.
For the $k$th fold, the observations with indices $i\in I_k$ are treated as test sites and the remaining sites are used as training data.
The proposed mixture of GEE model is fitted to $\{(\bm{y}_i,\bm{x}_i):i\notin I_k\}$, yielding estimates of the model parameters, including $\widehat{\bm{\alpha}}$, the mean parameters $\widehat{\mu}_g$, and the working correlation matrices $\widehat{\bm{R}}_g$.

For each test site $i\in I_k$, we use only the response components not contained in $J_k$ to infer its region membership.
Let $\bm{y}_{i,-J_k}$ denote the subvector of $\bm{y}_i$ corresponding to the components in $\{1,\ldots,p\}\setminus J_k$, and define $\mathcal{S}_{-J_k}(\bm{y}_i;\widehat{\bm{\theta}}_g)$ as the corresponding subvector of the estimating function.
Similarly, let $\widehat{\Omega}_{g,-J_k}^{-1}$ denote the covariance matrix of this reduced score.
The pseudo-probability of test site $i$ belonging to group $g$ is then defined as
\begin{align*}
\widehat{\tau}_{ig}^{(k)}
=
\frac{
\pi_{ig}(\widehat{\bm{\alpha}})
\phi_{p-|J_k|}\{\mathcal{S}_{-J_k}(\bm{y}_i;\widehat{\bm{\theta}}_g);
\bm{0}_{p - |J_k|},\widehat{\bm{\Omega}}_{g,-J_k}\}
}{
\sum_{g'=1}^G \pi_{ig'}(\widehat{\bm{\alpha}}) \phi_{p-|J_k|}\{
\mathcal{S}_{-J_k}(\bm{y}_i;\widehat{\bm{\theta}}_{g'});
\bm{0}_{p - |J_k|},\widehat{\bm{\Omega}}_{g',-J_k}\}
}.    
\end{align*}
Thus, the held-out response components in $J_k$ are not used when estimating the group-membership probabilities for the test sites.

For each held-out component $j\in J_k$, its predictive mean is then given by
$\widehat{\mu}_{ij}^{(k)}=\sum_{g=1}^G\widehat{\tau}_{ig}^{(k)}\widehat{\mu}_{gj}$.
For binary responses, we evaluate predictive performance using the Bernoulli log score,
$L_{ij}^{(k)}=y_{ij}\log\widehat{\mu}_{ij}^{(k)}
+(1-y_{ij})\log\{1-\widehat{\mu}_{ij}^{(k)}\}$.
For count responses, we instead use the Poisson log score,
$L_{ij}^{(k)}=\log p_{\rm Pois}(y_{ij};\widehat{\mu}_{ij}^{(k)})$.
The fold-specific predictive score is defined as
${\rm CV}_k(G)=\sum_{i\in I_k}\sum_{j\in J_k}L_{ij}^{(k)}$,
where a larger value indicates better predictive performance.

To avoid selecting an unnecessarily large number of groups, we employed the one-standard-error rule based on the cross validation criterion. 
Let $\overline{{\rm CV}}(G)=K^{-1}\sum_{k=1}^K {\rm CV}_k(G)$ denote the average cross-validation score for a candidate value of $G$, and let ${\rm SE}(G)$ denote its estimated standard error across folds.
Rather than simply choosing the value of $G$ that maximizes $\overline{{\rm CV}}(G)$, we employ the one-standard-error rule to favor a more parsimonious model.
Specifically, we let $G_{\max}=\arg\max_G\overline{{\rm CV}}(G)$, and then select
\[
\widehat{G}=
\min\left\{ G: \overline{{\rm CV}}(G) \geq \overline{{\rm CV}}(G_{\max}) - {\rm SE}(G_{\max})
\right\}.
\]
Hence, among the candidate models whose predictive performance is within one standard error of the best-performing model, we choose the one with the smallest number of groups.
This rule reduces the tendency of predictive criteria to select unnecessarily complex mixture models.

\section{Simulation Study}
\label{sec:sim}

\subsection{Performance of estimation and inference under binary response}
\label{sec:sim-binary-estimation}

We evaluate the finite-sample performance of the proposed GEE mixture through simulation studies based on correlated multivariate binary data.
For each subject $i=1,\ldots,n$, let $\bm{x}_i=(1,X_{i1},X_{i2},X_{i3})^\top$ denote a covariate vector, where $X_{i1},X_{i2},X_{i3}$ were independently generated from the uniform distribution on $(0,1)$.
Given $\bm{x}_i$, the latent group membership $g_i\in\{1,\ldots,G\}$ was sampled from a multinomial distribution with probabilities determined by a multinomial logistic model,
$$
\Pr(g_i=g| \bm{x}_i)=\pi_{ig}
=\frac{\exp(\bm{x}^\top_i\bm{\alpha}_g)}{\sum_{g'=1}^G \exp( \bm{x}_i^\top \bm{\alpha}_{g'})},
\qquad g=1,\ldots,G,
$$
where the first group was taken as the baseline category.
We set $G=3$ and specified the true parameter values as $\bm{\alpha}_2=(0, 2, -2, 0)$ and $\bm{\alpha}_3=(0, -2, 2, 0)$.
Conditional on the group membership $g_i=g$, the marginal success probabilities of the $p$ binary responses were specified by a mean vector $\bm{\mu}_g=(\mu_{1g},\ldots,\mu_{pg})^\top$ for $g = 1,\ldots,G$, where we adopted two scenarios for the true values, described as
\begin{align*}
({\rm Scenario\ 1}) \ \ \ &\mu_1 = \mu_1^{\ast}\equiv (\underbrace{0.2,\ldots,0.2}_{p/3},\underbrace{0.2,\ldots,0.2}_{p/3},\underbrace{0.8,\ldots,0.8}_{p/3})^\top, \\
&\mu_2 = \mu_2^{\ast}\equiv (\underbrace{0.2,\ldots,0.2}_{p/3}, \underbrace{0.8,\ldots,0.8}_{p/3}, \underbrace{0.2,\ldots,0.2}_{p/3})^\top, \\
&\mu_3 = \mu_3^{\ast}\equiv (\underbrace{0.8,\ldots,0.8}_{p/3}, \underbrace{0.2,\ldots,0.2}_{p/3}, \underbrace{0.2,\ldots,0.2}_{p/3})^\top, \\ 
({\rm Scenario\ 2}) \ \ \ & \mu_{gj}=\mu_{gj}^{\ast} + U(-0.1, 0.1), \quad g=1,\ldots,G, \quad j=1,\ldots,p. 
\end{align*}
To introduce within-subject dependence among the $p$ binary responses, we considered three types of working correlation structures, as follows:

\begin{itemize}
\item
(ID)\  All three groups have independent outcomes, that is, $\bm{R}_g=\bm{I}_p$ for $g=1,2,3$.

\item
(Mix1) \ The first group $(g=1)$ has an exchangeable correlation structure, $R_1=(1-\rho_1)\bm{I}_p+\rho_1 \bm{J}_p$ with $\rho_1=0.3$, where $\bm{J}_p$ denotes the $p\times p$ matrix of ones.
The second group $(g=2)$ has an AR(1) correlation structure, $(R_2)_{jk}=\rho_2^{|j-k|}$ with $\rho_2=0.7$.
For the third group $(g=3)$, an unstructured correlation matrix is generated by first drawing $\bm{\Sigma}_3\sim W_p(p+2,\bm{\Sigma}_0)$ with $\bm{\Sigma}_0=(1-\rho_0)\bm{I}_p+\rho_0\bm{J}_p$ and $\rho_0=0.1$, and then standardizing $\bm{\Sigma}_3$ to obtain the correlation matrix $R_3$.

\item
(Mix2)\ The three groups have the same types of correlation structures as in Mix1, but with $\rho_1=0.4$, $\rho_2=0.5$ and $\rho_0=0.2$.
\end{itemize}

For the simulated dataset, we applied the proposed MixGEE with two implementations of working correlations. 
The first method, denoted by ``MG1", employed an unstructured working correlation matrix and used the corresponding model-based covariance without a sandwich correction.
The second method, denoted by ``MG2", adopted an independence working correlation matrix but used a sandwich covariance estimator to account for possible within-subject dependence among the multivariate responses.
For comparison, we also considered an independence model, denoted by ``ID", which used an independence working correlation matrix without a sandwich correction.
Thus, MG2 differs from ID in that it retains robustness to within-subject correlation through the sandwich covariance estimator, even though both methods use an independence working correlation structure.
For comparison, we also fitted a region of common profiles model \citep{foster2013modelling} using the \texttt{ecomix} package \citep{skip2026ecomix} in R, denoted by ``RCP", as a likelihood-based benchmark for site-level clustering of multivariate species data.
Using 300 replications, we computed the mean squared errors of $\bm{\eta}$ (logit-transformed mean) and $\bm{\alpha}$ (coefficient of classification probability). 
For inference, we constructed pointwise 90\% confidence intervals for $\bm{\eta}$ using $B=100$ bootstrap replications. 
For MG1, MG2, and ID, we used the clustered Dirichlet random-weight bootstrap described in Section~\ref{sec:variance}, while for RCP we used the nonparametric bootstrap implemented in the \texttt{ecomix} package. 
We evaluated the performance through coverage probability (CP) and average interval length (AIL), where the results are shown in Tables~\ref{tab:sim-bin-mse} and \ref{tab:sim-bin-interval}.

When the responses are independent, the four methods generally exhibit comparable estimation accuracy for both $\bm{\eta}$ and $\bm{\alpha}$. 
The proposed methods MG1 and MG2, as well as ID, also attain coverage probabilities close to the nominal level. 
RCP tends to produce slightly shorter confidence intervals, accompanied by some undercoverage even under the independence setting.
The differences among the methods become more apparent when within-subject correlation is present. 
In terms of estimation of $\bm{\eta}$, accounting for the dependence structure through either the working correlation matrix (MG1) or the sandwich covariance estimator (MG2) generally improves performance relative to ID, particularly in the more strongly correlated settings. 
A similar pattern is observed for interval estimation: the coverage probabilities of ID deteriorate under Mix1 and Mix2, whereas MG1 and MG2 maintain substantially better coverage. 
These results indicate that incorporating within-subject dependence is particularly important for reliable uncertainty quantification, even when the marginal mean model itself is correctly specified.
For RCP, the MSE increases when the sample size increases from $n=500$ to $n=1000$. 
Inspection of the individual Monte Carlo replications suggests that this is driven by a small number of extremely poor estimates, possibly reflecting instability of the likelihood-based mixture fit, which may be further exacerbated when within-subject dependence is ignored.

\begin{table}[htb!]
\centering
\begin{tabular}{cccccccccccccccc}
\hline
&&& \multicolumn{4}{c}{MSE ($\bm{\eta}$)} && \multicolumn{4}{c}{MSE ($\bm{\alpha}$)}\\
\cline{4-7}\cline{9-12}
{\small Scenario} & $n$ & Cor & MG1 & MG2 & ID & RCP &  & MG1 & MG2 & ID & RCP \\
\hline
 &  & ID & 4.05 & 4.06 & 3.99 & 3.94 &  & 1.61 & 1.61 & 1.62 & 1.55 \\
1 & 500 & Mix1 & 5.75 & 5.70 & 10.6 & 5.09 &  & 1.81 & 1.74 & 1.98 & 1.77 \\
 &  & Mix2 & 6.55 & 5.51 & 7.75 & 4.61 &  & 1.69 & 1.70 & 1.69 & 1.62 \\
 \hline
 &  & ID & 1.95 & 1.95 & 1.93 & 9.23 &  & 0.53 & 0.53 & 0.53 & 0.48 \\
1 & 1000 & Mix1 & 3.30 & 2.98 & 3.76 & 19.6 &  & 0.58 & 0.55 & 0.65 & 0.62 \\
 &  & Mix2 & 5.93 & 4.79 & 4.03 & 33.5 &  & 0.58 & 0.59 & 0.59 & 0.57 \\
 \hline
 &  & ID & 4.42 & 4.42 & 4.58 & 4.32 &  & 1.70 & 1.76 & 1.63 & 1.56 \\
2 & 500 & Mix1 & 8.28 & 6.12 & 16.9 & 6.20 &  & 1.76 & 1.70 & 1.86 & 1.67 \\
 &  & Mix2 & 8.54 & 5.85 & 13.6 & 4.89 &  & 1.66 & 1.69 & 1.69 & 1.59 \\
 \hline
 &  & ID & 2.17 & 2.17 & 2.16 & 34.3 &  & 0.55 & 0.55 & 0.53 & 0.49 \\
2 & 1000 & Mix1 & 5.54 & 3.57 & 8.65 & 25.6 &  & 0.58 & 0.55 & 0.65 & 0.59 \\
 &  & Mix2 & 5.63 & 3.00 & 4.39 & 41.7 &  & 0.55 & 0.56 & 0.57 & 0.54 \\
\hline
\end{tabular}
\caption{Mean squared errors (MSEs) of the estimators of $\bm{\eta}$ (region-specific mean parameter) and $\bm{\alpha}$ (multinomial-logit coefficients) under binary response with two scenarios of marginal success probability and three underlying correlation structures, based on 300 Monte Carlo replications.
The MSE for $\bm{\eta}$ and $\bm{\alpha}$ are multiplied by 100 and 10, respectively. 
}
\label{tab:sim-bin-mse}
\end{table}

\begin{table}[htb!]
\centering
\begin{tabular}{cccccccccccccccc}
\hline
&&& \multicolumn{4}{c}{CP (\%)} && \multicolumn{4}{c}{AIL}\\
\cline{4-7}\cline{9-12}
{\small Scenario} & $n$ & Cor &   MG1 & MG2 & ID & RCP &  & MG1 & MG2 & ID & RCP \\
\hline
 &  & ID & 90.2 & 89.9 & 90.8 & 85.8 &  & 0.661 & 0.654 & 0.658 & 0.574 \\
1 & 500 & Mix1 & 89.4 & 86.9 & 91.4 & 83.7 &  & 0.789 & 0.679 & 0.978 & 0.611 \\
 &  & Mix2 & 90.9 & 86.9 & 92.4 & 85.9 &  & 0.925 & 0.670 & 0.969 & 0.618 \\
 \hline
 &  & ID & 90.2 & 90.2 & 90.3 & 85.0 &  & 0.461 & 0.461 & 0.459 & 0.404 \\
1 & 1000 & Mix1 & 87.9 & 89.7 & 87.8 & 79.9 &  & 0.664 & 0.502 & 0.585 & 0.475 \\
 &  & Mix2 & 92.7 & 88.1 & 91.1 & 84.3 &  & 0.842 & 0.491 & 0.632 & 0.473 \\
 \hline
 &  & ID & 90.5 & 90.1 & 91.2 & 86.0 &  & 0.689 & 0.677 & 0.707 & 0.595 \\
2 & 500 & Mix1 & 88.8 & 87.9 & 90.1 & 84.3 &  & 0.780 & 0.705 & 1.023 & 0.629 \\
 &  & Mix2 & 89.8 & 88.1 & 91.2 & 85.7 &  & 0.815 & 0.695 & 1.006 & 0.630 \\
 \hline
 &  & ID & 89.9 & 89.9 & 90.2 & 84.7 &  & 0.477 & 0.475 & 0.477 & 0.436 \\
2 & 1000 & Mix1 & 88.4 & 89.6 & 88.3 & 81.3 &  & 0.568 & 0.513 & 0.662 & 0.459 \\
 &  & Mix2 & 90.0 & 89.4 & 91.1 & 84.2 &  & 0.635 & 0.496 & 0.644 & 0.453 \\
\hline
\end{tabular}
\caption{Coverage probability (CP) and average interval length (AIL) of $90\%$ confidence intervals $\bm{\eta}$ (region-specific mean parameter) under binary response with two scenarios of marginal success probability and three underlying correlation structures, based on 300 Monte Carlo replications.}
\label{tab:sim-bin-interval}
\end{table}

\subsection{Performance of selecting the number of groups under binary response}\label{sec:sim-binary-selection}

We next evaluate the performance of selecting the number of latent groups under binary responses.
The data-generating settings are identical to those considered in Section~\ref{sec:sim-binary-estimation}.
The true number of groups is $G=3$ throughout this experiment.
For MG1, MG2, and ID, we considered candidate values $G=2,\ldots,7$ and selected the number of groups using the cross-validation criterion described in Section~\ref{sec:CV}.
For the competing likelihood-based method, denoted by RCP, the number of groups was selected by minimizing the BIC over the same range of candidate values.

Table~\ref{tab:sim-bin-select} reports the probability of correctly selecting the true number of groups, $\Pr(\widehat{G}=3)$, together with the average selected number of groups, based on 100 Monte Carlo replications.
When the outcomes were independent, all methods almost perfectly recovered the true number of groups for both sample sizes and both scenarios.
The differences among the methods became more pronounced when within-subject dependence was present.
Overall, MG1 showed relatively stable selection performance under both Mix1 and Mix2, and its performance generally improved as the sample size increased.
The corresponding average values of $\widehat{G}$ were also close to the true value.
In contrast, MG2 tended to overestimate the number of groups when the responses were correlated, particularly under Mix2.
This tendency became more pronounced for $n=1000$, where the average selected number of groups was close to five in Scenario 1 and above four and a half in Scenario 2.
Although ID performed well under Mix1, its selection accuracy deteriorated under the more challenging Mix2 setting, especially for larger sample sizes.
The RCP method exhibited a strong tendency to overestimate the number of groups whenever within-subject correlation was present, never selecting the true value in either Mix1 or Mix2.
These results suggest that explicitly estimating the within-group correlation structure, as in MG1, can substantially stabilize selection of the number of latent groups when multivariate binary outcomes exhibit heterogeneous dependence.

\begin{table}[htb!]
\centering
\begin{tabular}{ccccccccccccccc}
\hline
&&& \multicolumn{4}{c}{Probability of $\hat{G}=3$} && \multicolumn{4}{c}{Average of $\hat{G}$} \\
\cline{4-7}\cline{9-12}
{\small Scenario} & $n$ & Cor & MG1 & MG2 & ID & RCP &  & MG1 & MG2 & ID & RCP \\
\hline
&  & ID & 1.00 & 1.00 & 1.00 & 1.00 &  & 3.00 & 3.00 & 3.00 & 3.00 \\
1 & 500 & Mix1 & 0.74 & 0.52 & 0.89 & 0.00 &  & 3.43 & 3.71 & 3.15 & 4.96 \\
 &  & Mix2 & 0.68 & 0.16 & 0.77 & 0.00 &  & 3.47 & 4.36 & 3.30 & 5.00 \\
 \hline
 &  & ID & 1.00 & 1.00 & 1.00 & 0.99 &  & 3.00 & 3.00 & 3.00 & 3.01 \\
1 & 1000 & Mix1 & 0.91 & 0.36 & 0.95 & 0.00 &  & 3.09 & 4.16 & 3.05 & 6.51 \\
 &  & Mix2 & 0.87 & 0.05 & 0.63 & 0.00 &  & 3.24 & 4.92 & 3.61 & 5.91 \\
 \hline
 &  & ID & 1.00 & 0.99 & 1.00 & 1.00 &  & 3.00 & 3.01 & 3.00 & 3.00 \\
2 & 500 & Mix1 & 0.73 & 0.46 & 0.84 & 0.00 &  & 3.32 & 3.76 & 3.18 & 5.00 \\
 &  & Mix2 & 0.60 & 0.15 & 0.64 & 0.00 &  & 3.58 & 4.47 & 3.50 & 5.00 \\
 \hline
 &  & ID & 1.00 & 1.00 & 1.00 & 0.99 &  & 3.00 & 3.00 & 3.00 & 3.01 \\
2 & 1000 & Mix1 & 0.92 & 0.58 & 0.92 & 0.00 &  & 3.12 & 3.71 & 3.08 & 6.59 \\
 &  & Mix2 & 0.71 & 0.06 & 0.53 & 0.00 &  & 3.47 & 4.62 & 3.64 & 5.73 \\
\hline
\end{tabular}
\caption{Probability of selecting the true number of groups, $\hat{G}=3$, and average of $\hat{G}$, based on 100 Monte Carlo replications under binary response. }
\label{tab:sim-bin-select}
\end{table}

\subsection{Performance of estimation and inference under count response}\label{sec:sim-count}

We next evaluate the estimation and inference performance under multivariate count responses.
We use the same settings for the covariates, group-membership probabilities, sample sizes, response dimension, and types of region-specific correlation structures as in the binary-response experiment in Section~\ref{sec:sim-binary-estimation}.
In particular, we set $G=3$, $p=30$, and $n\in\{500,1000\}$, and consider the three correlation scenarios ID, Mix1, and Mix2 described in the previous subsection.
The ID scenario remains unchanged, while for Mix1 we set $(\rho_1,\rho_2,\rho_0)=(0.4,0.6,0.3)$ and for Mix2 we set $(\rho_1,\rho_2,\rho_0)=(0.6,0.9,0.4)$, using the same exchangeable, AR(1), and unstructured correlation structures for the three groups, respectively.

Conditional on group membership $g_i=g$, let $\bm{\mu}_g=(\mu_{1g},\ldots,\mu_{pg})^\top$ denote the corresponding mean vector.
We consider the following two scenarios for these mean parameters.
\begin{align*}
({\rm Scenario\ 1}) \ \ \ 
&\bm{\mu}_1 = \bm{\mu}_1^{\ast}\equiv
(\underbrace{4,\ldots,4}_{p/3},
 \underbrace{8,\ldots,8}_{p/3},
 \underbrace{12,\ldots,12}_{p/3})^\top, \\
&\bm{\mu}_2 = \bm{\mu}_2^{\ast}\equiv
(\underbrace{12,\ldots,12}_{p/3},
 \underbrace{8,\ldots,8}_{p/3},
 \underbrace{4,\ldots,4}_{p/3})^\top, \\
&\bm{\mu}_3 = \bm{\mu}_3^{\ast}\equiv
(\underbrace{8,\ldots,8}_{p/3},
 \underbrace{12,\ldots,12}_{p/3},
 \underbrace{8,\ldots,8}_{p/3})^\top, \\ 
({\rm Scenario\ 2}) \ \ \ 
&\mu_{gj}
=
\mu_{gj}^{\ast}
+
U(-2,2),
\qquad
g=1,\ldots,G,\quad j=1,\ldots,p.
\end{align*}
The perturbations in Scenario 2 are generated independently across groups and response components.
The parameter of interest on the link scale is $\eta_{jg}=\log(\mu_{jg})$.

Given $g_i=g$, we generate $\bm{Z}_i\sim\mathcal{N}_p(\bm{0}_p,\bm{R}_g)$ and set $U_{ij}=\Phi(Z_{ij})$ and $y_{ij}=F_{\mathrm{Pois}(\mu_{jg})}^{-1}(U_{ij})$ for $j=1,\ldots,p$, where $\Phi$ denotes the standard normal distribution function and $F_{\mathrm{Pois}(\mu)}^{-1}$ denotes the quantile function of a Poisson distribution with mean $\mu$.
This Gaussian copula construction ensures that $y_{ij}|g_i=g\sim\mathrm{Pois}(\mu_{jg})$ marginally, while inducing within-subject dependence through $\bm{R}_g$.

For each simulated dataset, we apply the same three versions of the proposed MixGEE approach as in the binary-response experiment.
MG1 uses the regularized unstructured working correlation matrix described in Section~\ref{sec:mixgee}, MG2 uses an independence working correlation matrix with the sandwich covariance estimator, and ID uses an independence working correlation matrix without the sandwich correction.
As a likelihood-based competitor, we fit the region of common profiles model using a Poisson specification, denoted by RCP.
The simulation is repeated 300 times.
We assess estimation accuracy using the mean squared errors of $\bm{\eta}$ and $\bm{\alpha}$.
For inference on $\bm{\eta}$, we construct pointwise 90\% confidence intervals using the same bootstrap procedures and number of bootstrap replications as in Section~\ref{sec:sim-binary-estimation}, and report their coverage probabilities (CP) and average interval lengths (AIL).

Tables~\ref{tab:sim-count-mse} and \ref{tab:sim-count-interval} report the corresponding results for count responses. 
Broadly, the results show patterns similar to those observed in the binary response experiment in Section~\ref{sec:sim-binary-estimation}. 
MG1 and MG2 provide stable estimation of the region-specific mean parameters across the different correlation structures, while the performance of methods that ignore within-subject dependence tends to deteriorate when correlation is present. 
In particular, the coverage probabilities of ID decrease noticeably under Mix1 and Mix2, whereas MG1 and MG2 remain considerably closer to the nominal level. The deterioration is even more pronounced for RCP in several settings.
An additional issue in this experiment is that the data generating distribution is not exactly Poisson. 
Thus, the Poisson likelihood assumed by RCP is misspecified even under the independence setting. This likelihood misspecification may partly explain the relatively poor performance of RCP, including the substantial undercoverage observed for $n=1000$. 
In contrast, the proposed MG approach does not require specification of the full marginal or joint distribution and therefore appears to be less sensitive to this form of distributional misspecification. 
When within-subject correlation is additionally present, the very low coverage probabilities of both ID and RCP further illustrate the importance of accounting for dependence among the multivariate responses.

\begin{table}[htb!]
\centering
\begin{tabular}{cccccccccccccccc}
\hline
&&& \multicolumn{4}{c}{MSE ($\bm{\eta}$)} && \multicolumn{4}{c}{MSE ($\bm{\alpha}$)}\\
\cline{4-7}\cline{9-12}
Scenario & $n$ & Cor &  MG1 & MG2 & ID & RCP &  & MG1 & MG2 & ID & RCP \\
\hline
 &  & ID & 0.09 & 0.09 & 0.09 & 299.0 &  & 4.88 & 5.10 & 4.13 & 57.8 \\
1 & 500 & Mix1 & 0.09 & 0.09 & 0.10 & 375.3 &  & 1.85 & 3.06 & 1.62 & 48.1 \\
 &  & Mix2 & 0.11 & 0.11 & 0.25 & 332.2 &  & 1.74 & 2.09 & 1.92 & 79.3 \\
 \hline
 &  & ID & 0.04 & 0.04 & 0.04 & 0.04 &  & 3.45 & 3.66 & 2.93 & 0.48 \\
1 & 1000 & Mix1 & 0.04 & 0.04 & 0.06 & 0.05 &  & 0.73 & 2.37 & 0.56 & 0.54 \\
 &  & Mix2 & 0.05 & 0.04 & 0.20 & 0.24 &  & 0.69 & 1.39 & 1.00 & 1.11 \\
 \hline
 &  & ID & 0.09 & 0.09 & 0.09 & 222.3 &  & 5.63 & 5.79 & 5.13 & 27.4 \\
2 & 500 & Mix1 & 0.09 & 0.09 & 0.10 & 201.7 &  & 4.32 & 4.97 & 1.63 & 31.9 \\
 &  & Mix2 & 0.09 & 0.09 & 0.19 & 158.7 &  & 3.37 & 3.57 & 1.79 & 63.7 \\
 \hline
 &  & ID & 0.04 & 0.04 & 0.04 & 0.04 &  & 5.12 & 5.22 & 4.59 & 0.48 \\
2 & 1000 & Mix1 & 0.04 & 0.04 & 0.05 & 0.05 &  & 3.27 & 4.30 & 0.56 & 0.51 \\
 &  & Mix2 & 0.04 & 0.04 & 0.10 & 0.13 &  & 2.81 & 3.28 & 0.73 & 0.79 \\
\hline
\end{tabular}
\caption{Mean squared errors (MSEs) of the estimators of $\bm{\eta}$ (region-specific mean parameter) and $\bm{\alpha}$ (multinomial-logit coefficients) under count response with two scenarios of marginal success probability and three underlying correlation structures, based on 300 Monte Carlo replications.
The MSE for $\bm{\eta}$ and $\bm{\alpha}$ are multiplied by 100 and 10, respectively. }
\label{tab:sim-count-mse}
\end{table}

\begin{table}[htb!]
\centering
\begin{tabular}{cccccccccccccccc}
\hline
&&& \multicolumn{4}{c}{CP (\%)} && \multicolumn{4}{c}{AIL}\\
\cline{4-7}\cline{9-12}
Scenario & $n$ & Cor &  MG1 & MG2 & ID & RCP &  & MG1 & MG2 & ID & RCP \\
\hline
 &  & ID & 89.2 & 89.2 & 89.2 & 18.8 &  & 0.093 & 0.093 & 0.093 & 0.065 \\
1 & 500 & Mix1 & 89.1 & 88.8 & 87.6 & 18.7 &  & 0.094 & 0.093 & 0.098 & 0.072 \\
 &  & Mix2 & 89.4 & 88.8 & 77.1 & 17.5 &  & 0.104 & 0.095 & 0.118 & 0.089 \\
  \hline
 &  & ID & 89.4 & 89.4 & 89.2 & 84.4 &  & 0.066 & 0.066 & 0.066 & 0.057 \\
1 & 1000 & Mix1 & 88.7 & 89.4 & 85.2 & 82.8 &  & 0.069 & 0.065 & 0.069 & 0.061 \\
 &  & Mix2 & 89.5 & 89.4 & 66.3 & 63.2 &  & 0.068 & 0.066 & 0.083 & 0.084 \\
  \hline
 &  & ID & 89.3 & 89.4 & 89.2 & 32.0 &  & 0.094 & 0.094 & 0.094 & 0.082 \\
2 & 500 & Mix1 & 88.9 & 88.8 & 88.0 & 40.5 &  & 0.094 & 0.094 & 0.097 & 0.096 \\
 &  & Mix2 & 89.0 & 88.9 & 83.0 & 39.9 &  & 0.098 & 0.095 & 0.108 & 0.115 \\
  \hline
 &  & ID & 89.5 & 89.4 & 89.6 & 84.6 &  & 0.067 & 0.067 & 0.067 & 0.058 \\
2 & 1000 & Mix1 & 89.8 & 89.6 & 88.9 & 84.8 &  & 0.067 & 0.067 & 0.068 & 0.060 \\
 &  & Mix2 & 89.7 & 90.1 & 80.7 & 75.6 &  & 0.070 & 0.067 & 0.076 & 0.071 \\
\hline
\end{tabular}
\caption{Coverage probability (CP) and average interval length (AIL) of $90\%$ confidence intervals $\bm{\eta}$ (region-specific mean parameter) under count response with two scenarios of marginal success probability and three underlying correlation structures, based on 300 Monte Carlo replications.}
\label{tab:sim-count-interval}
\end{table}

\section{Application to Kerguelen Plateau fish assemblage data}
\label{sec:app}

We applied the proposed method to multivariate presence--absence data for fish species collected at sampling sites around the Kerguelen Plateau.
The dataset consisted of $n=524$ sampling sites, with the response vector at each site comprising binary occurrence indicators for $p=20$ fish taxa.
We used eight environmental variables as site-level covariates: average depth, log-transformed slope, log-transformed current speed, sea-floor temperature, mean bottom nitrate concentration, mean temperature, annual variability in chlorophyll-a concentration, and variability in sea-surface-height anomaly.
All covariates were standardized to have mean zero and unit variance before fitting the models.

We first fitted the proposed MixGEE for binary responses, using a region-specific unstructured working correlation matrix.
We evaluated the cross-validation criterion with $K=10$ in Section~\ref{sec:CV} over $G\in\{2,\ldots,7\}$ and selected the number of groups using the one-standard-error rule.
This resulted in $G=3$.
For comparison, we also fitted the regions of common profiles (RCP) model \citep{foster2013modelling} using the \texttt{ecomix} package.
For RCP, the number of groups was selected by minimizing the Bayesian information criterion over $G\in\{2,\ldots,7\}$, resulting in $G=5$.
As shown in the simulation study in Section~\ref{sec:sim-binary-selection}, the BIC-based selection for RCP tends to overestimate the number of groups when substantial within-site correlation is present.
Since the present data also exhibit non-negligible dependence among taxa, as shown below, the selected value $G=5$ may partly reflect this tendency toward overestimation.

\begin{figure}
\centering
\includegraphics[width=\linewidth]{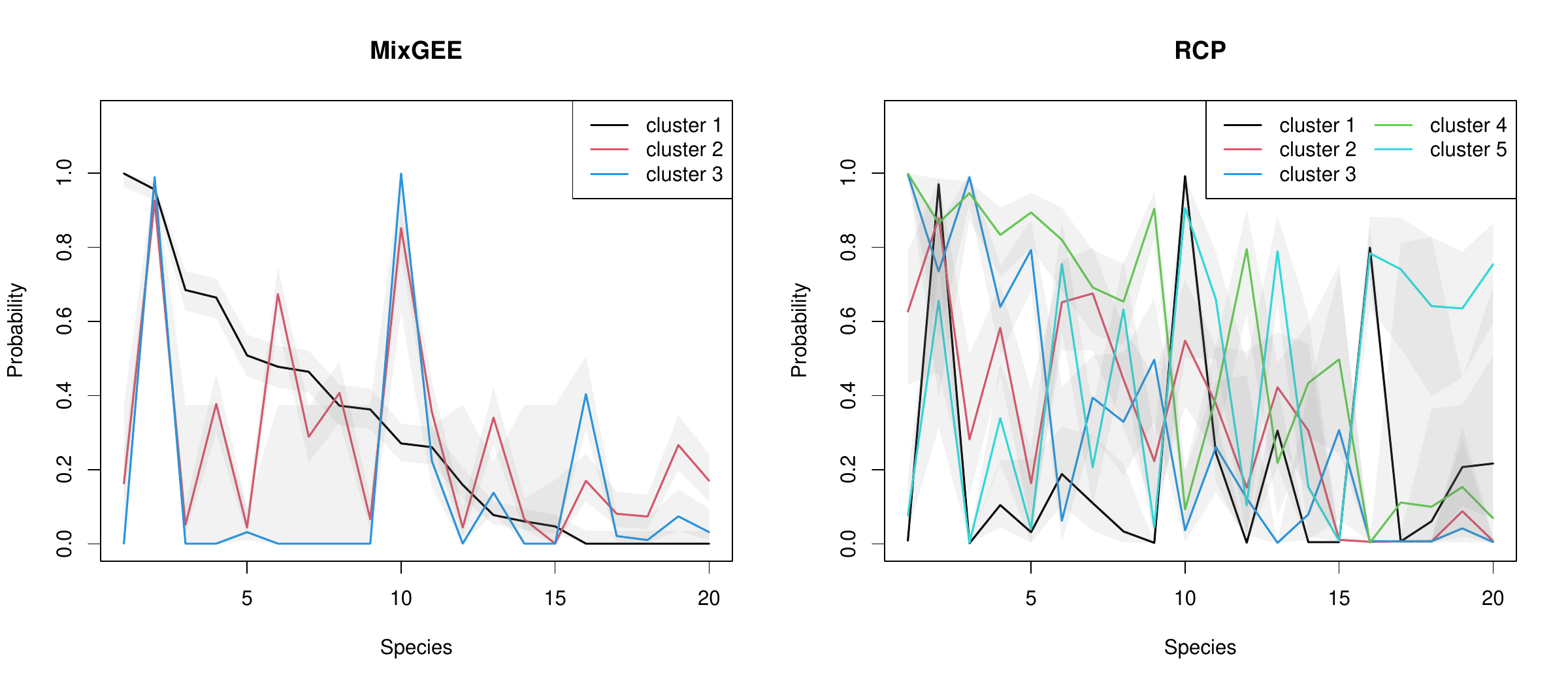}
\caption{
Estimated region-specific occurrence probabilities for the 20 fish taxa
obtained from the proposed MixGEE and RCP models.
Shaded regions represent pointwise 90\% confidence intervals.
}
\label{fig:app-eta}
\end{figure}

Using the selected number of groups, we estimated the region-specific occurrence probabilities for both methods. 
Following the bootstrap procedures used in the simulation study, pointwise 90\% confidence intervals were constructed using $B=100$ bootstrap replications for both MixGEE and RCP. 
Including these bootstrap replications, the computation times were 12.2 seconds for MixGEE and 7.3 seconds for RCP on the same computing environment, indicating that the two methods had comparable computational costs in this application.
Figure~\ref{fig:app-eta} presents the estimated region-specific occurrence probabilities for the 20 taxa under MixGEE and RCP.
For ease of comparison, the taxa are ordered according to their estimated occurrence probabilities in the largest group under MixGEE.
The estimated occurrence profiles show substantial heterogeneity across groups: several taxa have moderate or high occurrence probabilities in one group while having much smaller probabilities in the other groups.
Thus, the estimated groups represent distinct patterns of community composition rather than merely differences in overall species prevalence.

The RCP fit yields a finer partition with five groups, compared with the three groups identified by MixGEE.
Given the simulation results on group-number selection and the substantial within-site dependence observed in this dataset, this finer partition should be interpreted with some caution.
In particular, part of the additional splitting under RCP may arise from dependence among taxa that is explicitly accommodated by the working correlation structure in MixGEE.

We next compare the spatial classifications produced by MixGEE and RCP.
For each observed site, we assign the site to the group with the largest estimated covariate-dependent mixing probability.
The same rule can also be applied to unobserved locations because the mixing probabilities depend only on the environmental covariates.
We therefore evaluated the fitted mixing probabilities over a regular spatial grid covering the study region and assigned each grid location to its most likely group.

\begin{figure}
\centering
\includegraphics[width=\linewidth]{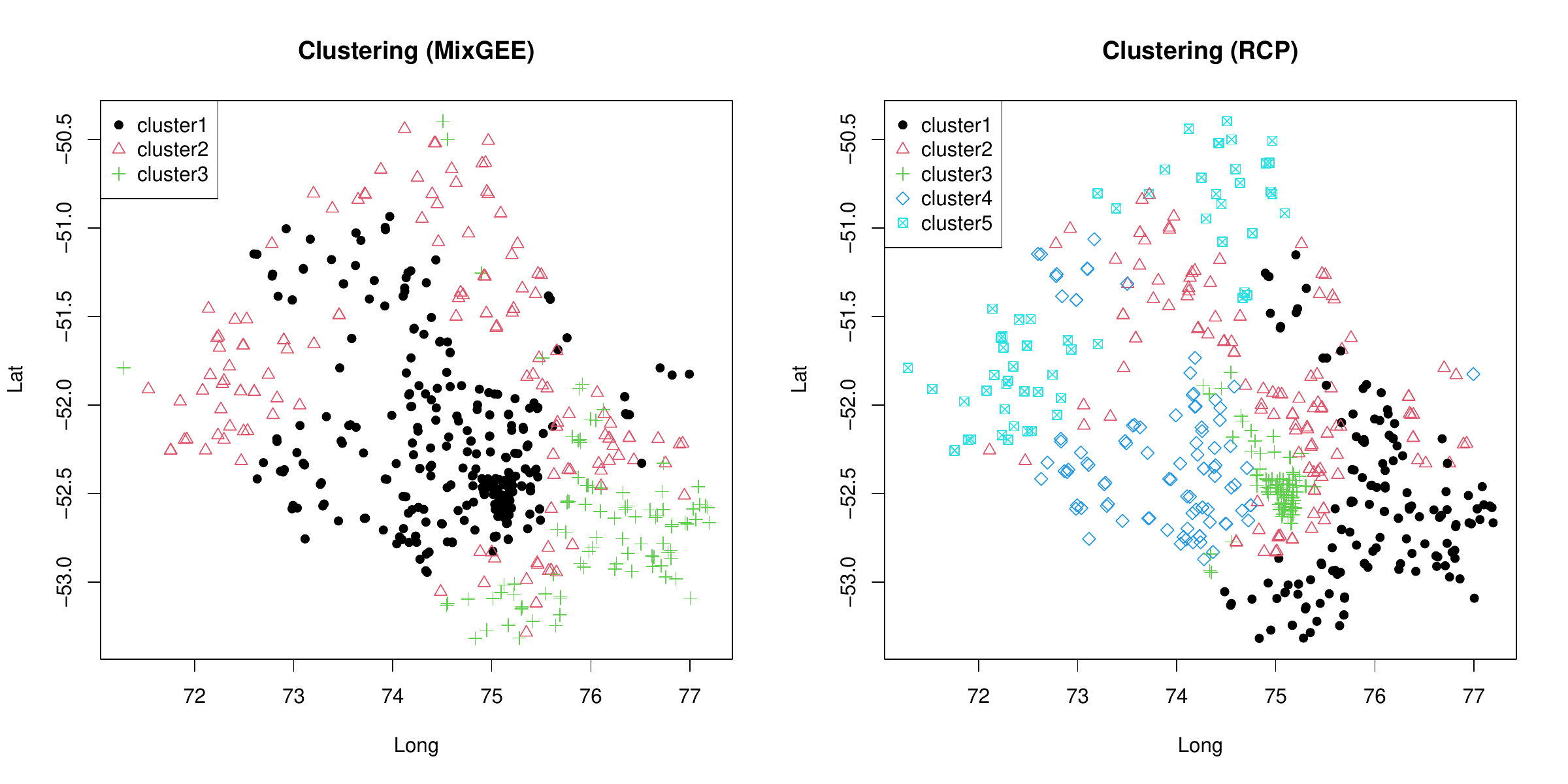}
\includegraphics[width=\linewidth]{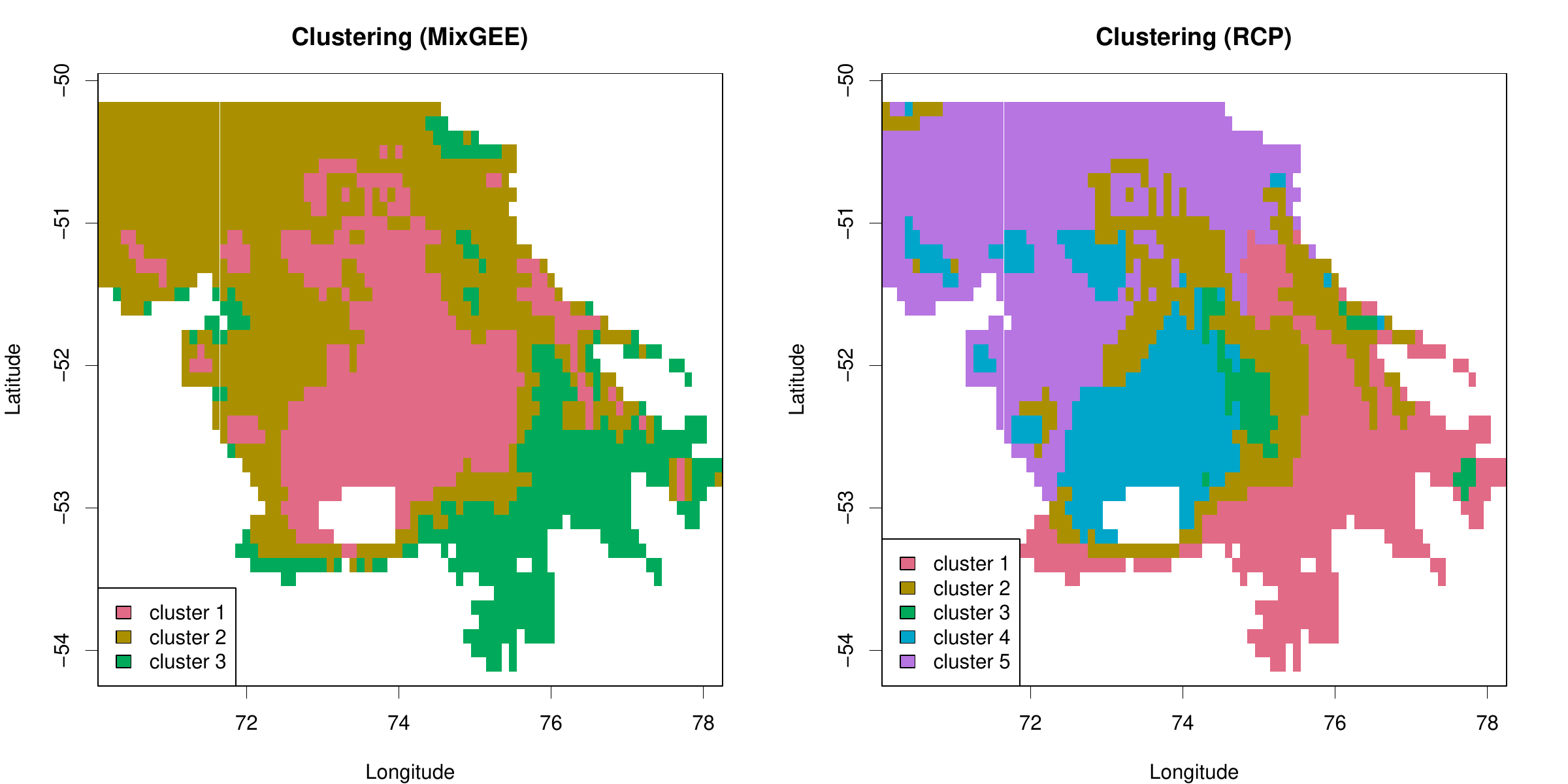}
\caption{
Spatial classifications obtained from MixGEE and RCP for the observed sites
(upper panels) and unobserved locations on a regular spatial grid
(lower panels).
Each location is assigned to the group with the largest estimated
covariate-dependent mixing probability.
}
\label{fig:app-cluster}
\end{figure}

Figure~\ref{fig:app-cluster} shows the resulting classifications for the observed sites and the regular spatial grid.
Both methods produce spatially structured group assignments, indicating that the environmental covariates induce systematic geographical variation in community composition.
At the same time, the RCP classification is divided into a larger number of spatial regions, consistently with its selection of $G=5$, whereas MixGEE gives a more parsimonious classification with three groups.
The spatial predictions also illustrate an advantage of modeling the mixing probabilities as functions of environmental covariates: the fitted model provides a direct classification of locations at which species observations are unavailable.

\begin{figure}
\centering
\includegraphics[width=\linewidth]{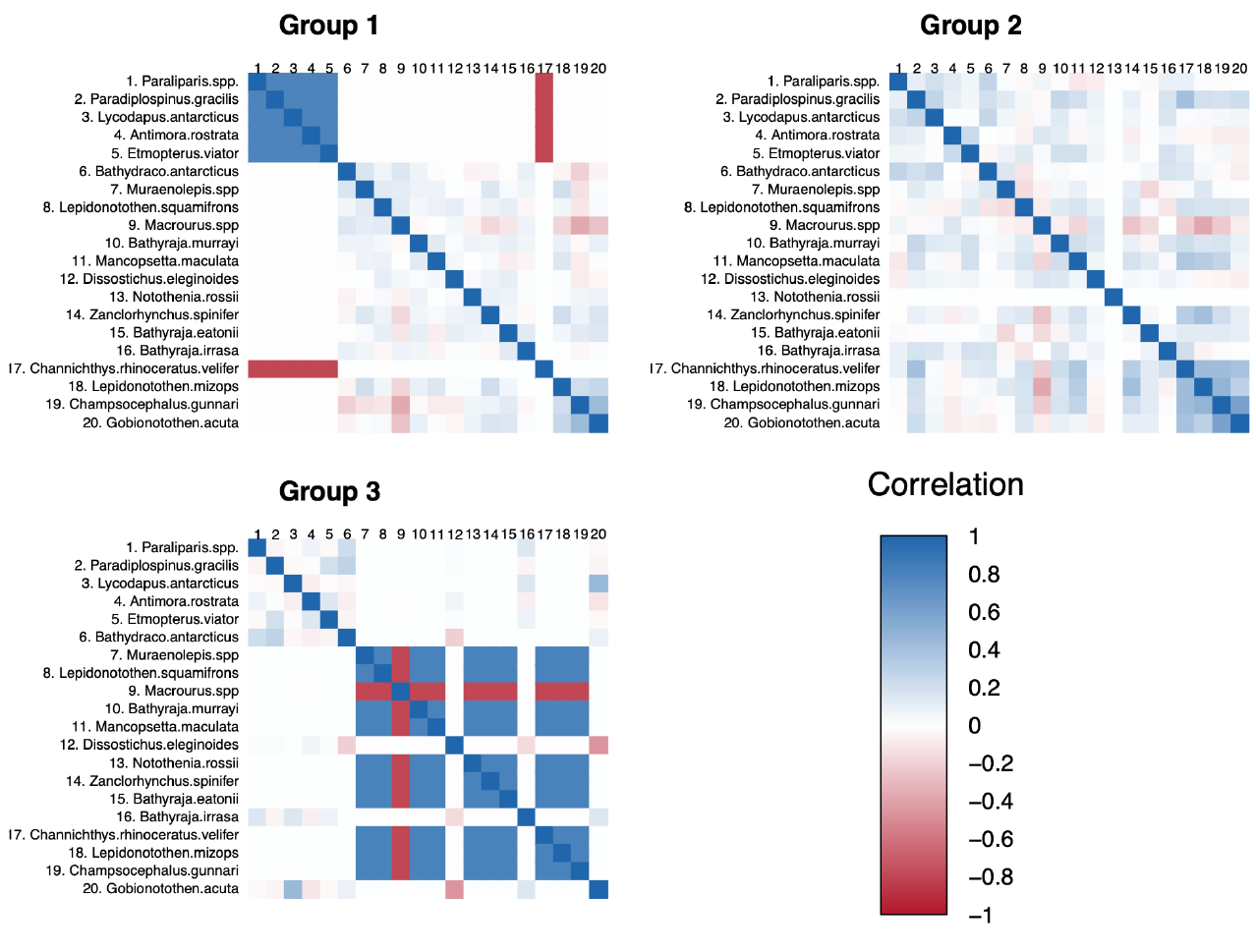}
\caption{
Estimated region-specific unstructured working correlation matrices for
the three groups identified by the proposed MixGEE method.
}
\label{fig:app-cor}
\end{figure}

Figure~\ref{fig:app-cor} displays the estimated region-specific unstructured working correlation matrices for the three groups, obtained by MixGEE.
The estimated correlation patterns differ considerably across groups and include both positive and negative associations among taxa.
This suggests that the dependence structure among species is itself heterogeneous across the latent groups and cannot be adequately represented by a single common correlation matrix.

This feature also highlights an important distinction between MixGEE and RCP.
While both methods allow the marginal occurrence profiles to vary across groups, MixGEE additionally accommodates region-specific within-site dependence
through the working correlation matrices.
The application therefore illustrates that heterogeneity in multivariate presence--absence data can arise not only through differences in marginal occurrence probabilities but also through differences in the association structure among taxa.
Accounting for both sources of heterogeneity leads to a more parsimonious three-group representation of the fish assemblage data in this application.

\begin{figure}
\centering
\includegraphics[width=\linewidth]{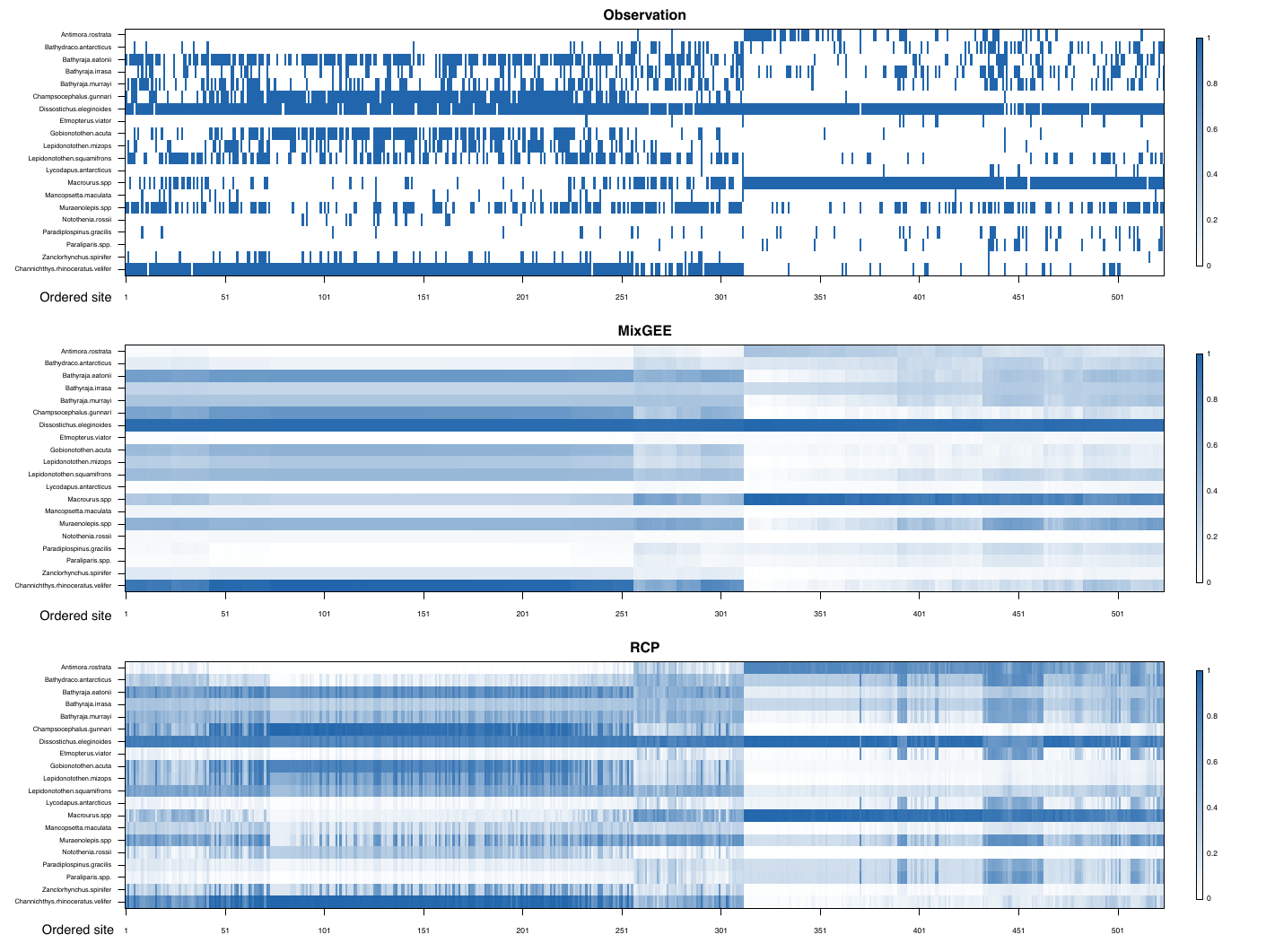}
\caption{
Observed presence--absence indicators for the 20 fish taxa (top) and
estimated site-specific occurrence probabilities under MixGEE (middle)
and RCP (bottom). Sites are ordered according to hierarchical clustering
of the occurrence probability profiles estimated by MixGEE, and the same
ordering is used in all panels. Darker colors indicate observed
occurrence in the top panel and higher estimated occurrence probabilities
in the middle and bottom panels.
}
\label{fig:app-occur}
\end{figure}

To further compare the fitted community patterns under MixGEE and RCP, we computed the site-specific occurrence probabilities by averaging the region-specific occurrence probabilities with respect to the estimated covariate-dependent mixing probabilities. 
Figure~\ref{fig:app-occur} compares the observed presence--absence patterns with the fitted site-specific occurrence probabilities obtained from MixGEE and RCP. 
Sites are ordered using hierarchical clustering of the occurrence probability profiles estimated by MixGEE, and the same ordering is used in all three panels.
MixGEE exhibits more clearly separated occurrence patterns across sites, whereas the RCP fit shows more gradual variation.
This difference can also be confirmed from the entropy of the estimated mixing probabilities.
Specifically, letting $\widehat{\pi}_{ig}$ denote the estimated mixing probability for site $i$ and group $g$, we compute the normalized entropy $-(n\log G)^{-1}\sum_{i=1}^n\sum_{g=1}^G\widehat{\pi}_{ig}\log\widehat{\pi}_{ig}$, which gives $0.394$ for MixGEE and $0.449$ for RCP.
This indicates that the estimated group memberships under MixGEE are more concentrated and hence more clearly separated across sites.
Although this difference cannot be attributed solely to the modeling of within-site dependence, it may partly reflect the use of region-specific working correlation structures in MixGEE, which allows information on associations among taxa to contribute to the identification of the latent groups.

Taken together, these results highlight an important distinction between MixGEE and RCP.
While both methods allow the marginal occurrence profiles to vary across groups, MixGEE additionally accommodates region-specific within-site dependence through the working correlation matrices.
The application therefore suggests that accounting for both marginal heterogeneity and dependence among taxa can lead to a more parsimonious and clearly separated representation of fish assemblage structure.

\section{Discussion}\label{sec:conc}

In this paper, we proposed a mixture modeling framework based on generalized estimating equations for correlated multivariate outcomes. The method accounts for within-observation dependence through working correlation structures without requiring a fully specified multivariate likelihood. We also developed a cross-validation procedure for selecting the number of latent groups. Simulation studies and the fish assemblage application demonstrated the practical usefulness of the proposed approach.

While the unstructured working correlation was considered in this paper, other parsimonious structures, such as exchangeable or AR(1) working correlations, can also be accommodated e.g., 
if $\bm{R}_g$ is characterized by a small number of region-specific working correlation parameters $\bm{\rho}_g$, then these can be estimated by minimizing $\|\widehat{\bm{R}}_g - \bm{R}_g(\bm{\rho}_g)\|_F^2$, where $\|\cdot\|_F$ denotes the Frobenius norm.

Several directions for further development remain. First, the present implementation models the marginal means through a $p$-dimensional vector of region-specific mean parameters that is constant across observations within each latent group. 
This specification was motivated by the clustering applications considered here and provides a particularly simple interpretation of the resulting group profiles. 
However, the framework can be readily extended to conventional GEE-type regression models in which the marginal mean of each response component depends on observation-specific covariates. 
The pseudo-posterior construction and iterative estimation scheme remain conceptually unchanged, allowing the proposed framework to accommodate within-group regression relationships in addition to latent clustering.
Second, the performance of the method depends to some extent on the choice and estimation of the working correlation structure. 
Although the sandwich formulation provides robustness against correlation misspecification, estimating a fully unstructured correlation matrix may become unstable when the response dimension is large relative to the number of observations within a group. 
More structured or sophisticated regularization of the working correlation matrix would therefore be useful in higher-dimensional settings \citep[e.g.][]{warton2011regularized}. 
Extensions in these directions would broaden the applicability of the proposed mixture GEE framework while retaining its main advantage of avoiding a fully specified multivariate likelihood.

\section*{Acknowledgments}
Shonosuke Sugasawa was supported by JSPS KAKENHI Grant Numbers 24K21420 and 25H00546. Francis Hui was supported by an Australian Research Council Discovery Project DP240100143.

\bibliographystyle{apalike}
\bibliography{ref}

\newpage
\appendix
\section{Additional details for estimating MixGEE} \label{app:mixgeedetails}

In this section, we here provide some explicit forms relevant when applying MixGEE to the case where $y_{ij}$ correspond to binary and count responses, respectively. These two cases differ in their mean--variance relationships, but can be handled within the same mixture GEE framework described above. Both response types were considered in the simulation study in Section \ref{sec:sim} of the main text.

\subsubsection*{Binary responses}
We specify $\mathbb{E}[y_{ij}| z_{ig} = 1] = \mu_{gj}$, use the inverse logit link function $\mu_{gj} = \{1 + \exp(\eta_{gj})\}^{-1}\exp(\eta_{gj})$, and assume a quadratic marginal variance function ${\rm Var}[y_{ij}| z_{ig} = 1] = \mu_{gj}(1-\mu_{gj})$. It follows that $\bm{A}_g = \text{Diag}\{\mu_{g1}(1-\mu_{g1}),\ldots,\mu_{gp}(1-\mu_{gp})\}$ and $\bm{D}_g = \bm{A}_g$, such that equation \eqref{eq:EE-y} in the main text simplifies with $\mathcal{S}(\bm{y}_i; \bm{\eta}_g) = \bm{A}_g^{1/2} \bm{R}_g^{-1}\bm{A}_g^{-1/2} (\bm{y}_i - \bm{\mu}_g)$. Moreover, since $\bm{A}_g$ does not depend on $i$, then the GEE reduces even further to the solution $\widehat{\bm{\mu}}_g = (\sum_{i=1}^n z_{ig})^{-1} \sum_{i=1}^n z_{ig} \bm{y}_i$, with the corresponding estimate on the link scale obtained component-wise as
$\widehat{\eta}_{gj} = \log\{(1-\widehat{\mu}_{gj})^{-1} \widehat{\mu}_{gj}\}$. Step iii in Algorithm \ref{algo:EEE} uses these updates with $z_{ig}$ replaced by $\tau^{(t+1)}_{ig}$ as appropriate.

\subsubsection*{Count responses}
For count responses, a standard approach is to adopt a Poisson mean--variance type relationship, such that $\mathbb{E}[y_{ij}| z_{ig} = 1] = \mu_{gj}$ with the exponential link function $\mu_{gj} = \exp(\eta_{gj})$ and the variance function ${\rm Var}[y_{ij}| z_{ig} = 1] = \mu_{gj}$. It follows that $\bm{A}_g = \text{Diag}(\mu_{g1},\ldots,\mu_{gp})$ and $\bm{D}_g = \bm{A}_g$, such that equation \eqref{eq:EE-y} in the main text again simplifies with $\mathcal{S}(\bm{y}_i; \bm{\eta}_g) = \bm{A}_g^{1/2} \bm{R}_g^{-1}\bm{A}_g^{-1/2} (\bm{y}_i - \bm{\mu}_g)$. Since $\bm{A}_g$ does not depend on $i$, then the GEE reduces even further to the solution $\widehat{\bm{\mu}}_g = (\sum_{i=1}^n z_{ig})^{-1} \sum_{i=1}^n z_{ig} \bm{y}_i$, with the corresponding estimate on the link scale obtained component-wise as
$\widehat{\eta}_{gj} = \log(\widehat{\mu}_{gj})$. Step iii in Algorithm \ref{algo:EEE} uses these updates with $z_{ig}$ replaced by $\tau^{(t+1)}_{ig}$ as appropriate.

Thus, although binary and count responses have different variance functions and link functions, the update of the marginal means for each region has the same simple weighted-average form.

\end{document}